\documentclass[12pt,letterpaper]{article}
\pdfoutput=1 
\usepackage{amssymb}
\usepackage{amsmath}
\usepackage{amsfonts}
\usepackage{epsfig}
\usepackage{feynmf}
\usepackage{color}
\newcommand{\sect}[1]{\setcounter{equation}{0}\section{#1}}

\usepackage{bm}
 \font\small=cmr10 scaled \magstep0
  
\outer\def\beginsection#1\par{\medbreak\bigskip
      \message{#1}\leftline{\bf#1}\nobreak\medskip
\vskip-\parskip
      \noindent}

\newcommand{\eq}{\begin{equation}}
\newcommand{\eqx}{\end{equation}}
\newcommand{\eqn}{\begin{eqnarray}}
\newcommand{\eqnx}{\end{eqnarray}}
\newcommand{\bi}{\begin{itemize}}
\newcommand{\ei}{\end{itemize}}

\usepackage{color}

\def\be{\begin{equation}}
\def\ee{\end{equation}}
\def\ba{\begin{eqnarray}}
\def\ea{\end{eqnarray}}

\begin{document}
\begin{titlepage}
\hfill \hbox{CERN-TH-2017-186}
\vskip 0.1cm
\hfill \hbox{NORDITA-2017-093}
\vskip 0.1cm
\hfill \hbox{TAUP-3023/17}
\vskip 0.5cm
\begin{flushright}
\end{flushright}
\begin{center}
{\Large \bf Spontaneous $CP$ breaking in QCD and the axion potential: 
 an effective Lagrangian approach}
 \vskip 1.0cm {\large Paolo
Di Vecchia$^{a, b}$, Giancarlo Rossi$^{c,d}$,
Gabriele Veneziano$^{e,f}$, \\
Shimon Yankielowicz$^{g}$ } \\[0.7cm]
{\it $^a$ The Niels Bohr Institute, Blegdamsvej 17, DK-2100 Copenhagen {\O}, Denmark}\\
{\it $^b$ Nordita, KTH Royal Institute of Technology and Stockholm 
University, \\Roslagstullsbacken 23, SE-10691 Stockholm, Sweden}\\
{\it $^c$ Dipartimento di Fisica, Universit\`a di Roma ``Tor Vergata'' and INFN Sezione 
Roma 2, \\ Via della Ricerca Scientifica - 00133 Roma, Italy}\\
{\it $^{d}$ Centro Fermi - Museo Storico della Fisica e Centro Studi e Ricerche ``E.\ Fermi''}\\
{\it Piazza del Viminale 1 - 00184 Roma, Italy}\\
{\it $^e$ Coll\`ege de France, 11 place M. Berthelot, 75005 Paris, France}\\
{\it $^f$Theory Division, CERN, CH-1211 Geneva 23, Switzerland}\\
{\it $^g$ School of Physics and Astronomy, Tel-Aviv University, Ramat-Aviv 69978 Israel}\\
\end{center}
\begin{abstract}
Using the well-known  low-energy effective Lagrangian of QCD --valid for small (non-vanishing) quark masses and a large number of colors-- we study in detail the regions of parameter space where $CP$ is spontaneously broken/unbroken for a vacuum angle $\theta=  \pi$. In the $CP$ broken region there are first order phase transitions as one crosses $\theta=\pi$, while on the (hyper)surface separating the two regions, there are second order phase transitions signalled by the vanishing of the mass of a pseudo Nambu-Goldstone boson and by a divergent QCD topological susceptibility. The second order point sits at the end of a first order line associated with the $CP$ spontaneous breaking, in the appropriate complex parameter plane. When the effective Lagrangian is extended by the inclusion of an axion these features of QCD imply that standard calculations of the axion potential have to be revised if the QCD parameters fall in the above mentioned $CP$ broken region, in spite of the fact that the axion solves the strong-$CP$ problem. These last results could be of interest for axionic dark matter calculations if the topological susceptibility of pure Yang-Mills theory falls off sufficiently fast when temperature is increased towards the QCD deconfining transition.

 \end{abstract}

\end{titlepage}

\sect{Introduction}
\label{intro}
\setcounter{equation}{0}

Already in the early seventies Dashen recognized~\cite{RD} that phases in the quark mass matrix could spontaneously break $CP$ and the possibility that such a phenomenon could explain the observed  $CP$ violation in kaon physics was explored~\cite{JN}. It turned out that these violations were too large to explain the experiments with $K$ mesons and would give a much too high value for the electric dipole moment of the neutron and for the $\eta \rightarrow 2 \pi$ decay amplitude~\cite{MB}. 
At about the same time Weinberg pointed out~\cite{SWCP} that possible  $CP$ violating phases can be eliminated through chiral rotations of the quark fields. These rotations included an anomalous $U_A(1)$ transformation and therefore generated a $CP$ violating term proportional to $F \tilde{F}$.
However, at the time such a term was considered innocuous since it amounts to adding to the Lagrangian a total derivative (and, indeed, it is irrelevant at all orders in perturbation theory). It looked therefore as if QCD did automatically conserve $CP$. 

The phenomenological problem with that naive conclusion is that the same triviality 
of $F \tilde{F}$ implies the famous $U(1)$ problem, expressed for instance by the anomalously large $\eta'$ mass.
After the discovery of the instanton solutions  and  the presence of
different topological  sectors in pure Yang-Mills (YM) theory, it was soon realized~\cite{tHooft} that the $U(1)$ problem might be solved although this remained somewhat controversial for a while~\cite{RC}. The observation~\cite{EW1,GV} that, in the framework of large-$N$ 
QCD, the mass matrix of the mesons contains, besides the
terms related to the masses of the quark, an extra parameter connected to 
the topological susceptibility of pure YM theory, opened 
the way to  a quantitative resolution of the $U(1)$ 
problem~\cite{DiVecchia:1981aev,Giusti:2001xh,DelDebbio:2004ns,Ce:2015qha}~\footnote{A big role in the solution of the $U(1)$ problem was played by the analogy of QCD with the $CP^{n-1}$ model in two dimensions~\cite{PDV1}.}.

Unfortunately, the resolution of the $U(1)$ problem brought back the question of $CP$ conservation in strong interactions.
Indeed, $CP$ violating phases of the quark mass matrix could no longer be rotated away so that QCD would not automatically preserve $CP$.
The YM Lagrangian could be supplemented with an extra term,
given by  the  topological charge density  and containing a parameter, the so-called vacuum angle  $\theta$, 
that also breaks $CP$. By performing an anomalous $U_A(1)$ transformation of the quark fields,  it turns out that 
the  relevant observable quantity  is a  combination of the $\theta$ parameter 
and the  phases present in the quark mass matrix $M$, given by $\bar{\theta} \equiv \theta + \arg \det m$. 
The $CP$ violation induced by a non-vanishing  $\bar{\theta}$ was
first used to estimate the resulting electric dipole moment of the neutron in~\cite{VB}.
It was later refined in~\cite{Crewther:1979pi} by identifying a leading logarithmic contribution thus establishing  a limit on  $\bar{\theta}$ of order $10^{-9}-10^{-10}$ for the smallness of which QCD, on its own, has no explanation. In Sect.~\ref{axion} we will come back to this problem and to its resolution with the help of an axion.
 
The next step was the construction and study of an extension~\cite{RST,DVV,NA,EW} 
of the effective Lagrangian of the light
pseudo Nambu-Goldstone bosons (the non-linear $\sigma$-model) to include a term linear  in 
the topological charge density and reproducing both the $U_A(1)$ anomaly and the $\theta$ 
term of the microscopic theory, as well as  a quadratic term whose coefficient is associated with 
the topological susceptibility of pure YM theory~\footnote{Together with Refs.~\cite{RST,DVV,NA,EW} see also Refs.~\cite{KO,NO} and Refs.~\cite{PDV,DVS} for an old and a more recent review.}.

The $\theta$ dependence of physical quantities, in the framework of the effective
Lagrangian for mesons, was studied in detail in Refs.~\cite{DVV,EW}
where it was found that for a generic non-zero value of $\theta$ $CP$ is broken
but, for $\theta=\pi$ (where $CP$ is a symmetry of the theory) could be either spontaneously broken or independent of 
the values of the quark masses and the topological susceptibility.

The possibility of spontaneously breaking of $CP$  from the introduction of phases in the quark mass matrix
was taken up again in~\cite{Smilga:1998dh,MC1,MC2} in the framework of low-energy effective 
Lagrangian for the pseudoscalar mesons, where it was shown that at $\theta = \pi$  there 
are indeed  two regions in parameter space, one where $CP$ is conserved and the other 
where $CP$ is broken, separated by a surface whose shape depends on the quark mass 
ratios. An important result of the analysis of Ref.~\cite{MC1} is that, on the separating 
surface, one of the mesons becomes massless.
  
Recently, the discussion of the case $\theta=\pi$ has been taken up again
in a very interesting paper~\cite{GKKS} where it was proven, under a few very plausible assumptions, that, even  for finite $N$, $CP$ must be spontaneously broken at $\theta=\pi$ in $SU(N)$ YM theory. The main ingredient in the derivation of this result is the use of 't Hooft's anomaly constraint for the mixed anomaly of the discrete $CP$ and center symmetries. This first order transition nicely fits with the spontaneous $CP$ breaking in QCD at $\theta=\pi$ in the decoupling (heavy quark mass) limit.

In the first part of this paper we discuss again the $\theta$ dependence of chiral, large-
$N$ QCD in its low-energy approximation, using the above mentioned effective
Lagrangian and concentrating our attention on what happens in the neighborhood of $\theta = \pi$.  Besides the quark masses, parametrized in terms of the $N_f$ parameters $-2 m_i \langle \bar{\psi} \psi \rangle \equiv \mu_i^2 F_{\pi}^2$,  there is an additional parameter,  the topological susceptibility of YM theory, $\chi_{YM}$, which, as already mentioned, plays a crucial role in the large-$N$ resolution of the $U(1)$ problem.
In this enlarged parameter space (w.r.t. the one considered in~\cite{MC1}) there is an hypersurface separating the region  where $CP$ is conserved from the one where $CP$
is spontaneously broken. On the hypersurface itself the theory exhibits a second order phase transition where one of the pseudo Nambu-Goldstone bosons (PNGBs) becomes exactly massless and 
the topological susceptibility of QCD diverges. Inside the $CP$ broken region the ground state makes a sudden, finite jump as $\bar{\theta}$  goes from $ \pi-\epsilon$ to 
$\pi+\epsilon$ corresponding to a first order phase transition. In an  appropriate complex parameter space (discussed in Sect.~\ref{QCDPD}) 
the second order point resides at the endpoint of a first order line associated with $CP$ breaking and starting at $-\infty$ 
where the decoupling to YM occurs. The position of the second order end-point resides  depends on all the other parameters (mass ratios, topological susceptibility). 

These results can be seen as a rather straightforward generalization of those of~\cite{MC1,MC2} to the case of a generic value of $\chi_{YM}$  and of~\cite{Nati,GKS} to the case of a generic quark mass matrix (the equal mass case is indeed quite special since it is always in the $CP$ broken phase except in the case of a single light flavor). In~\cite{GKS} the issue of $CP$ breaking in QCD was also addressed for finite $N$, and the theories residing on the resulting domain walls were studied.

In the second part of this paper we turn our attention to the case in which QCD has been augmented by the addition of an axion field, the best known way to solve, in a natural way, the strong-$CP$ problem. The axion can be easily incorporated in the effective Lagrangian (see e.g.~\cite{DVS}).
We then find that the QCD results of the previous Sections have an interesting bearing on the properties of the axion potential near the boundary of its periodicity interval.
Depending again on where one is in the QCD parameter  space the axion potential can differ significantly from the one commonly used in the literature (see e.g.~\cite{diCortona:2015ldu}). Furthermore, in the immediate vicinity of the critical hypersurface the very concept of an axion potential ceases to be physically meaningful since the dynamics is described by {\it two} very light pseudoscalars whose mass is of the order of the geometric mean between the PNGB mass and the conventional axion mass. Quite naturally, in that region the mass eigenstates are strongly mixed combinations of the two.
Although at zero temperature real QCD is quite deeply inside the $CP$ conserving region, one cannot exclude a-priori the possibility that, as one moves towards the deconfining, chiral-symmetry-restoring temperature, QCD may move (in parameter space) towards the critical hypersurface or even inside the $CP$ breaking region. If true, this could have interesting physical effects, e.g.\ on the standard computation of axionic dark matter abundance. As we will discuss, some precise lattice calculations in quenched QCD at finite temperature would be highly desirable in order to settle this point.

The paper is organized as follows. In Sect.~\ref{effective} we review the main
properties and consequences of the low-energy effective Lagrangian at generic values of the $\theta$ angle and quark masses.
In Sect.~\ref{Nf=1} we study in detail the behavior at $\theta=\pi$ in the case of a single flavor, while in Sects.~\ref{Nf=2} and~\ref{Nf} we discuss the case of two or more flavors respectively. Non-trivial checks that the results derived from the effective Lagrangian exactly satisfy general Ward-Takahshi identities (WTIs) are presented in Appendix~\ref{appA}. In Sect.~\ref{axion} we consider QCD with a very generic additional axionic degree of freedom and discuss the axion potential in the different situations described above. In particular we examine the "realistic" case of two or three unequal mass light flavors. Some final remarks are presented in Sect.~\ref{conclusions}.
  
\section{Chiral, large-$N$ QCD at arbitrary $\theta$: a reminder}
\label{effective}
\setcounter{equation}{0}

For the sake of being self-contained we summarize in this section some already known facts. We will refer, where appropriate, to the original literature for further details.

Assuming confinement and spontaneous  chiral symmetry breaking by a quark-antiquark condensate at a generic value of  $\theta$, QCD, for three light quarks  ($m_i \ll \Lambda_{QCD}$) and a large number of colors ($N \gg1$)~\footnote{We also assume to be below the so-called conformal window whose beginning is expected to occur at a value of $N_f$ proportional to $N$. Using the two-loop beta function it is found to occur at $N_f=34 N^3/(13 N^2-3)\sim \frac{34}{13} N$.},  is 
described at low-energy by the following effective Lagrangian~\cite{RST,DVV,NA,EW}
\begin{eqnarray}
&&L = \frac{1}{2} {\rm Tr} \left( \partial_\mu U \partial^\mu U^{\dagger} 
\right)  +
 \frac{F_\pi}{2\sqrt{2}} {\rm Tr} \left[ \mu^2 (U+U^\dagger) \right]   \nonumber \\
 &&+ 
\frac{Q^2}{2 \chi_{YM}}  
  + \frac{i}{2} Q  {\rm Tr} \left[ \log U - \log U^\dagger 
\right] - \theta Q \, .
\label{InitialL3}
\end{eqnarray}
Here $F_\pi$ is the pion decay constant  ($F_\pi \sim 95 MeV$ in the real world with $N=3$)~\footnote{Remember that $F_\pi$ grows like $\sqrt{N}$ for large $N$.} and
the  $3\times 3$ matrix $U$ describes, non-linearly,  the spontaneous breaking of the approximate $U(3)_L\otimes U(3)_R$ chiral  symmetry in terms of  nine light PNGBs so that 
\begin{eqnarray}
U = \frac{F_\pi}{\sqrt{2}}  {\rm e}^{i \sqrt{2} \Phi /F_\pi}~~;~~~ 
\Phi = \Pi^{a}  T^{a}_{ij} \, ,
\label{UN2}
\end{eqnarray}
where $T^{a}_{ij}$ are the matrices satisfying the algebra of $U(3)$ normalized 
as ${\rm Tr}(T^a T^b ) = \delta^{ab}$.  Furthermore, $\mu^2$ is proportional to the quark mass matrix~\footnote{In the literature $\mu^2$ is often denoted by $M$. In this paper we prefer this different notation in order to avoid confusion with a different use of the symbol $M$.} which, without loss
of generality, can be taken to be real, diagonal and non negative (provided a $\theta$-term is added). More precisely, in terms of the quark masses $m_i$ and condensate at $\theta =0$, $ \langle \bar{\psi} \psi \rangle$, $\mu^2$ is defined by
\begin{eqnarray}
\mu_{ij}^2 = \mu_i^2 \delta_{ij} = - 2 m_i  \langle \bar{\psi} \psi \rangle F_\pi^{-2}  \delta_{ij}\, .
\label{Mij}
\end{eqnarray}

Although the physically relevant  case is the one with two or three light flavors, for the
sake of generality,  we will consider hereafter the case of $N_f$ light flavors (hence now
$i,j=1, \dots, N_f$). $Q$ is the QCD topological charge density that appears in the divergence of the $U_A(1)$ current 
\begin{eqnarray}
&&\partial_{\mu} J^{\mu}_5 = 2 N_f Q + 2\sum_{i=1}^{N_f} m_i P_i
~~;~~~ Q = \frac{g^2}{32 \pi^2} F_{\mu \nu}^a
({ \tilde{F}}^a)^{\mu \nu}~~;~~ ({\tilde{F}}^{\mu \nu})^a = \frac{1}{2} \epsilon^{\mu \nu \rho \sigma}
F_{\rho \sigma}^a \nonumber \\
&& J^\mu_5 = \sum_{i=1}^{N_f} {\bar{\psi}}_i \gamma^\mu \gamma_5 \psi_i ~~;~~
P_i = i {\bar{\psi}}_i \gamma_5 \psi_i \, .
\label{Q+anomaly}
\end{eqnarray}

Modulo the mass term,  the Lagrangian~(\ref{InitialL3}) is invariant under $SU(N_f)_L \otimes SU(N_f )_R \otimes U(1)_V$ transformations while,  under the $U_A(1)$  transformation $U \rightarrow U {\rm e}^{-2i \alpha}$, one has
\begin{eqnarray}
\frac{i}{2} {\rm Tr } \left( \log U - \log U^\dagger \right) \rightarrow 
\frac{i}{2} {\rm Tr } \left( \log U - \log U^\dagger \right) + 2 \alpha N_f Q \, ,
\label{anomaly}
\end{eqnarray}
as needed. The quadratic term in $Q$ contains a
coefficient, $\chi_{YM}$, which  turns out to be nothing but the topological susceptibility of pure YM theory in the large-$N$ limit.  Finally, the last term takes into account of the presence of a non-zero $\theta$ parameter.

The  $2 \pi$ periodicity in $\theta$ (which in the underlying QCD theory is related to the quantization of $\nu \equiv \int d^4 x Q(x)$) can be easily checked at the level of~(\ref{InitialL3}). Indeed, a shift in $\theta$ by $2 \pi$ can be reabsorbed, thanks to the anomaly term in~(\ref{InitialL3}), by a chiral rotation by $2 \pi$ of a component (say $U_{11}$) of $U$ under which even the mass term in~(\ref{InitialL3}) is  invariant.
We also note that, under $CP$, $Q \rightarrow - Q$ and $U \rightarrow U^{\dagger}$. Thus naively, in our convention of real positive quark masses, only the last term in~(\ref{InitialL3}) breaks $CP$ unless $\theta =0$~\footnote{It is believed, and supported by lattice calculations and the chiral Lagrangian approach, that at $\theta=0$ the vacuum is non-degenerate and the theory is gapped with no spontaneous $CP$ breaking.}. However, even if $\theta = \pm \pi$, $CP$ is not {\it explicitly} broken since $2 \pi$ periodicity implies that $\theta = + \pi$ and $\theta = - \pi$ are equivalent. Nonetheless, as discussed below, $CP$ can be {\it spontaneously} broken at $\theta = \pm \pi$.

In the infinite-$N$ limit the anomaly effectively turns off and the physical PNGB spectrum consists of $N_f^2$ unmixed states of mass
\be
\label{PSmasses}
M^2_{ij} = \frac12( \mu_i^2 + \mu_j^2)  ~~,~~  i, j = 1, 2, \dots, N_f \, .
\ee
In general, one could add to the previous Lagrangian a $U(N_f)_L \otimes U(N_f )_R$ invariant  function of $Q$, $U$ and $U^\dagger$. However, it can be shown~\cite{RST,DVV,NA,EW} that 
the only  surviving terms at large $N$ are those appearing  in~(\ref{InitialL3}). 

Before we proceed further let us notice that the Lagrangian~(\ref{InitialL3}) for a single flavor is exactly the Lagrangian one gets by using the two-dimensional bosonization rules in the massive Schwinger model, where the  kinetic term of the gauge field corresponds to  the first term in the second line of~(\ref{InitialL3}) with $a \equiv \frac{e^2}{\pi}, F_\pi = \frac{1}{\sqrt{2\pi}}$, while the term coupling the fermions to the gauge field corresponds to the anomaly term with the logarithm. The other terms are also reproduced as also noticed in Ref.~\cite{MC2}. A similar structure appears also in other two-dimensional models as the one discussed in Ref.~\cite{Komargodski:2017dmc}. In those models, as also in the massive Schwinger model, the bosonized Lagrangian is  equivalent to the original microscopic Lagrangian, while, in our case, the effective Lagrangian~(\ref{InitialL3})
is only valid at low energy, for small quark masses, and for large $N$. However, the fact that in all these cases one gets the same Lagrangian indicates that our results may  not necessarily be valid only at large $N$. 

Since the equation of motion of $Q(x)$ is algebraic, we could integrate out $Q(x)$ 
from the start. However, as later on we will want  to compute the  $\langle Q Q \rangle$ correlator, we prefer to rewrite Eq.~(\ref{InitialL3}) as follows:
\begin{eqnarray}
&& L=   \frac{1}{2} {\rm Tr} \left( \partial_\mu U \partial^\mu U^{\dagger} 
\right)  +
 \frac{F_\pi}{2\sqrt{2}} {\rm Tr} \left( \mu^2 (U+U^\dagger) \right)  - \frac{\chi_{YM}}{2} 
 \left[ \theta - \frac{i}{2} {\rm Tr} \left( \log U - \log U^\dagger \right) \right]^2   \nonumber \\
&& + \frac{1}{2 \chi_{YM}} \left[ Q - \chi_{YM} \left( \theta - \frac{i}{2} {\rm Tr} \left( \log U
- \log U^\dagger \right)  \right)\right]^2 \, .
\label{ELLE}
\end{eqnarray}

The presence of the $\theta$ term  implies that, for unequal  masses,
 the vacuum does not correspond anymore
to $\langle U \rangle$ being  proportional  to the  unit matrix~\footnote{In spite of appearance, this does not correspond to a spontaneous breaking of $SU(N_f)_V$ since phases can always be rotated away into the quark mass matrix.}. We are obliged to introduce 
a separate VEV for each flavor by writing
\begin{eqnarray}
\langle   \Phi_{ij }\rangle = - \frac{F_\pi }{\sqrt{2}} \phi_i \delta_{ij} \, .
\label{vi}
\end{eqnarray}
Inserting Eq.~(\ref{vi}) in the previous Lagrangian the vacua of the theory correspond to the  minima of  the following  potential
\begin{eqnarray}
V (\phi_i ) = - \frac{F_\pi^2}{2} \sum_{i=1}^{N_f}  \mu_i^2 \cos \phi_i +
\frac{\chi_{YM}}{2} \left(\theta - \sum_{i=1}^{N_f} \phi_i  \right)^2 \, ,
\label{Vphiiphi}
\end{eqnarray} 
and are therefore obtained by looking for the stable solutions of the  equations
\begin{eqnarray}
&&  \mu_i^2 \sin \phi_i -  a 
\left(\theta - \sum_{j=1}^{N_f}  \phi_j \right)=0 ~~~;~~~i=1, \dots, N_f  \, ,
\label{minimum}
\end{eqnarray}
where we have defined 
\begin{eqnarray}
 a = \frac{2 \chi_{YM}}{F_\pi^{2}} \, .
 \label{defa}
\end{eqnarray}
The Eqs.~(\ref{minimum}) determine $\phi_i$ and all physical quantities in terms of $\mu_i^2, a$ and $\theta$.
Denoting this solution by $\phi_i = \hat{\phi}_i (\mu_i^2, a, \theta)$, and computing   
$\langle Q \rangle$ from the quadratic part of the Lagrangian in~(\ref{ELLE1}), 
we finally identify $\langle Q \rangle$  with  
$\chi_{YM} \left( \theta - \sum  \hat{\phi}_i \right)$.

Defining a new $\hat{U}$ matrix in terms of the shifted fields
\be
\hat{U}  \equiv  \frac{F_\pi}{\sqrt{2}}
{\rm e}^{ i \frac{\sqrt{2}}{F_\pi} \hat{\Phi}} ~~;~~ \hat{\Phi} = \Phi - \langle \Phi \rangle\, ,
\label{Uhat}
\ee
as well a shifted $Q$ field
 \be
\hat{Q}  =  Q  - \chi_{YM} ( \theta- \sum_{i=1}^{N_f}\hat{\phi}_i)\, ,
\label{Qhat}
\ee
we get a Lagrangian that depends on $\hat{U}$ and $\hat{Q}$ as follows 
\begin{eqnarray}
&& L = - V(\hat{\phi}_i) +    \frac{1}{2} {\rm Tr} \left( \partial_\mu \hat{U} \partial^\mu \hat{U}^{\dagger} 
\right)  + \frac{F_\pi^2}{2} {\rm Tr} \left[ \mu^2(\theta) \left( \cos \left(\frac{\sqrt{2}}{F_\pi} \hat{\Phi}
\right) -1\right) \right] - \frac{a}{2} \left[  {\rm Tr} \left(  \hat{\Phi} \right) \right]^2
\nonumber \\
&& + \chi_{YM}  (\theta - \sum_{i=1}^{N_f}\hat{\phi}_i )  {\rm Tr} \left[\sin 
\left( \frac{\sqrt{2}}{F_\pi}  \hat{\Phi} \right) - \frac{\sqrt{2}}{F_\pi} \hat{\Phi}   \right]
\nonumber \\
&& + \frac{1}{2 \chi_{YM}} \left[ \hat{Q} 
- \chi_{YM} \frac{\sqrt{2}}{F_\pi}  {\rm Tr} \,  \hat{\Phi} \right]^2\, ,
\label{ELLE1}
\end{eqnarray}
where we have defined
\begin{eqnarray}
\mu^2_{ij} (\theta) \equiv \mu_i^2 \cos \hat{\phi}_i \delta_{ij} \, .
\label{Mijtheta}
\end{eqnarray}
The first line of Eq.~(\ref{ELLE1}) (apart from the first term which is a constant)
describes the spectrum and the interaction 
of the PNGBs, the second, being odd under $ \hat{\Phi} \rightarrow -  \hat{\Phi}$,  gives the $CP$ violating contributions (controlled by its coefficient $ \chi_{YM} (\theta - \sum_{j=1}^{N_f} \hat{\phi}_i )$),
and the third line will be useful to determine the topological susceptibility in QCD. As we shall see below, while for $\theta =0$ the $CP$ violating coefficient is zero, for $\theta = \pm \pi$ it can be non-zero. The latter case has to be attributed to the spontaneous breaking of $CP$ by some non-$CP$-invariant VEVs.

The spectrum of the PNGBs is obtained by restricting our attention to the terms quadratic in $\hat{\Phi}$, coming from the first line of~(\ref{ELLE1}), for which we get
\begin{eqnarray}
L_2 = \frac{1}{2} {\rm Tr} \left( \partial_\mu \hat{\Phi} \partial^\mu \hat{\Phi} \right) - \frac{1}{2} {\rm Tr}
\left( \mu^2(\theta)\hat{\Phi}^2 \right) - \frac{a}{2} \left[  {\rm Tr} \left( \hat{\Phi}\right) \right]^2 \, .
\label{ELLE2}
\end{eqnarray}
Separating in  $\hat{\Phi}$ the generators in the Cartan sub-algebra from the others
\begin{eqnarray}
\hat{\Phi} = {\tilde{T}}^{\alpha \beta}_{ij} {\tilde{\Pi}}^{\alpha \beta} + v_i \delta_{ij}  \, ,
\label{Cartan1}
\end{eqnarray}
we have from $L_2$ the following two-point correlation functions in momentum space
\begin{eqnarray}
\langle {\tilde{\Pi}}^{\alpha \beta} (x) {\tilde{\Pi}}^{\gamma \delta} (y) \rangle^{F. T.} =
\frac{i \delta^{\alpha \gamma} \delta^{\beta \delta}}{ p^2 - M_{\alpha \beta}^2} ~~;~~
M_{\alpha \beta}^2 = \frac{1}{2} ( \mu_{\alpha}^2 (\theta) + \mu_{\beta}^2 (\theta) ) 
\label{noCartan}
\end{eqnarray}
and
\begin{eqnarray}
\langle v_i (x) v_{j} (y) \rangle^{F.T.} = i A^{-1}_{ij} (p^2) \, ,
\label{vivj}
\end{eqnarray}
where
\begin{eqnarray}
A_{ij} (p^2 ) = (p^2 - \mu_i^2 ) \delta_{ij} -a H_{ij} \equiv p^2 \delta_{ij} - M^2_{ij}
\label{Aij}
\end{eqnarray}
and $H_{ij}$ is a matrix with $1$ in all entries. 
The masses $M_i$  of the physical states in the Cartan sub-algebra  
are obtained by diagonalizing the matrix $ M^2_{ij}$ and  satisfy the  equation 
\begin{eqnarray}
\det  M^2 = \prod_{i=1}^{N_f} M_i^2 (\theta) = 
\prod_{i=1}^{N_f} \mu_i^2 (\theta)  \left[ 1 + a
\sum_{i=1}^{N_f} \frac{1}{\mu_i^2 (\theta) } \right] \, .
\label{detAp2=0}
\end{eqnarray}
For $p^2 \ne 0$ one gets
\begin{eqnarray}
\det A = \prod_{i=1}^{N_f} (p^2 - M_i^2 (\theta)) = 
\prod_{i=1}^{N_f} (p^2 - \mu_i^2 (\theta) ) \left[ 1 - a
\sum_{i=1}^{N_f} \frac{1}{p^2 - \mu_i^2 (\theta) } \right] \, .
\label{detA}
\end{eqnarray}
In the last part of this section we  use the Lagrangian~(\ref{ELLE1}) to compute the two-point correlator  of $\hat{Q}$ (note that, by definition $\langle \hat{Q} \rangle = \langle v_i\rangle = 0$) and relate the topological susceptibilities of YM and QCD. 
Since there is no quadratic term involving $v_i$ with the combination of $\hat{Q}$ and $v_j$ appearing in the last line of Eq.~(\ref{ELLE1}), we get immediately the following two-point correlation function 
\begin{eqnarray}
\langle \left( \hat{Q}(x) - \chi_{YM} \frac{\sqrt{2}}{F_\pi} \sum_{k=1}^{N_f} v_k (x) \right) 
v_j (y) \rangle =0 \, ,
\label{two1}
\end{eqnarray}
which implies
\begin{eqnarray}
\langle \hat{Q}(x) v_j (y) \rangle^{F.T.} = \frac{\chi_{YM} \sqrt{2}}{F_\pi}
 \sum_{k=1}^{N_f}  \langle v_k (x) v_j (y)  \rangle^{F.T.} = i \frac{\chi_{YM} \sqrt{2}}{F_\pi}
\sum_{k=1}^{N_f} A^{-1}_{kj} (p^2) \, ,
\label{Qv}
\end{eqnarray}
where in the last step we have used Eq.~(\ref{vivj}). From the relation
\begin{eqnarray}
\sum_{k=1}^{N_f} A^{-1}_{kj} (p^2 )= 
\frac{1}{p^2 - \mu_j^2 (\theta) } \frac{\prod_{i=1}^{N_f} (p^2 - \mu_i^2 (\theta))}{
\prod_{i=1}^{N_f} (p^2 - M_i^2 (\theta))}
\label{relut6}
\end{eqnarray}
the correlator~(\ref{Qv}) becomes 
\begin{eqnarray}
\langle \hat{Q}(x) v_j (y) \rangle^{F.T.} =i  \frac{\chi_{YM} \sqrt{2}}{F_\pi} 
\frac{1}{p^2 - \mu_j^2 (\theta)} \frac{\prod_{i=1}^{N_f} (p^2 - \mu_i^2 (\theta))}{
\prod_{i=1}^{N_f} (p^2 - M_i^2 (\theta))} \, .
\label{Qvj}
\end{eqnarray}
Finally, from the last line of Eq.~(\ref{ELLE1}) we get
\begin{eqnarray}
\langle \left( \hat{Q}  - \chi_{YM} \frac{\sqrt{2}}{F_\pi} \sum_{j=1}^{N_f} v_j  \right)\!\!(x) 
 \times \left( \hat{Q}   - \chi_{YM} \frac{\sqrt{2}}{F_\pi} \sum_{j=1}^{N_f} v_j  \right)\!\!(y) \rangle  =  i \chi_{YM} \delta^{(4)} (x-y) \, .
\label{(Q-<Q>)2}
\end{eqnarray}
Using Eq.~(\ref{Qvj})  and 
\begin{eqnarray}
 \sum_{h,k =1}^{N_f} A^{-1}_{hk} (p^2)  = \frac{ \sum_{k=1}^{N_f} \frac{1}{p^2 - \mu_k^2(\theta)}}
 {1 - a \sum_{k=1}^{N_f} \frac{1}{p^2 - \mu_k^2(\theta)}} = 
  \sum_{k=1}^{N_f} \frac{1}{p^2 - \mu_k^2(\theta)}
  \frac{\prod_{i=1}^{N_f} (p^2 - \mu_i^2 (\theta))}{
\prod_{i=1}^{N_f} (p^2 - M_i^2 (\theta))} \, ,
\label{equality}
\end{eqnarray}
we get
\begin{eqnarray}
\langle Q(x)  Q(y) \rangle_{conn.}^{F.T.} = \langle \hat{Q}(x) \hat{Q}(y)   \rangle^{F.T.} = 
 i  \frac{ \chi_{YM} }{1 - a \sum_{k=1}^{N_f} \frac{1}{p^2 - \mu_k^2 (\theta)}} \, .
\label{QQfinal}
\end{eqnarray}

In particular, for $p^2=0$ one gets the topological susceptibility in QCD with $N_f$ flavors 
\begin{eqnarray}
\chi_{QCD} = \frac{\chi_{YM}}{1 + a \sum_{i=1}^{N_f} \frac{1}{\mu_i^2 (\theta)} } = \chi_{YM}\left(1 - \frac{\chi_{YM}}{\sum_{k=1}^{N_f} ( m_i  \langle \bar{\psi} \psi \rangle)} \right)^{-1}\, .
\label{chiQCD}
\end{eqnarray}
Since our effective Lagrangian is, strictly speaking, valid for $N  \rightarrow \infty$ (where the $\eta'$ is a PNGB), the  quark condensate in the previous equation should be evaluated  in the leading planar order proportional to $N$. The next to the leading terms should not be included. In particular, it means that the next to the leading contributions which are affected by logarithmic 
divergencies~\cite{Sharpe:1992ft,Bernard:1992mk,Giusti:2002rx} 
and make the quenched quark condensate ill-defined, are avoided~\footnote{Even though our analysis is valid for large $N$, in the spirit of the large $N$ expansion, we will use it for the physical $N=3$ with the hope that even for this value the leading term (in the large $N$  expansion), dominates.}.

Finally as a last remark we wish to stress an important property of both Eqs.~(\ref{detAp2=0}) and~(\ref{chiQCD}), namely that they both reduce to the case of a theory 
with $N_f -1$ flavors when one of the quark
masses becomes very large. If all quarks become much heavier than $a$ (which can still be the case in the chiral regime since $a$ scales like $1/N$ at  large $N$) then 
$\chi_{QCD} \rightarrow \chi_{YM}$. Finally, when any quark flavor becomes massless the QCD topological susceptibility goes to zero as it should on general grounds.

In Appendix~\ref{appA} we provide the form of various two-point functions at small (but not necessarily vanishing) momenta and show that they satisfy exactly (i.e.\ without O$(1/N)$ 
corrections) all the expected anomalous and non-anomalous Ward--Takahashi identities (WTIs).

\section{QCD phase diagrams}
\label{QCDPD}
\setcounter{equation}{0}

In this section we discuss the phase diagrams of QCD at zero temperature and chemical potential  for different numbers of quark flavors $N_f$. The parameter space in which we consider possible phase transitions is
spanned by the $(N_f +1)$ parameters $\mu_i^2 \ge 0$ and $\theta$ (with $0 \le \theta < 2 \pi$) while considering $\chi_{YM}$ and $F_{\pi}$ (and thus $a$) as given.
In  Sect.~\ref{axion} we will see how those phase diagrams acquire a different meaning in the presence of a QCD axion and also briefly mention possible non-zero temperature effects.

Just to make our terminology clear. We will be talking about $CP$ conservation or violation referring, respectively, to the vanishing or non-vanishing of the quantity $ \chi_{YM} (\theta - \sum_{j=1}^{N_f} \phi_j )$ in Eq.~(\ref{ELLE1}). Sometimes the breaking of $CP$ is explicit 
(e.g.\ for generic values of $\theta$) while in some other cases it is spontaneous (like for $\theta = \pi$). We will try to make the distinction when needed in order to avoid confusion.

\subsection{$N_f =1$}
\label{Nf=1}

In the case of a single flavor the potential in Eq.~(\ref{Vphiiphi}) becomes, up to an irrelevant factor
\begin{eqnarray}
\frac{V (\phi)}{a} 
  = - \epsilon \cos \phi + \frac12 (\theta - \phi)^2~~;~~ \epsilon \equiv \frac{\mu^2}{a} \, ,
\label{potential}
\end{eqnarray}
from which we can compute its derivatives with respect to $\phi$
\begin{eqnarray}
\frac{V' }{a} &=&  \epsilon \sin \phi + \phi - \theta ~~;~~
\frac{V''}{a} =  \epsilon \cos \phi + 1 \nonumber  \\
\frac{V'''}{a} &=& - \epsilon \sin \phi ~~;~~  \frac{V''''}{a} = - \epsilon \cos \phi \, .
\label{potentialder}
\end{eqnarray}
Let us distinguish two cases:

\begin{itemize}

\item{$\epsilon < 1$} 

In this case $V'' >0$ so that there can only be a single stable minimum with positive mass. This is confirmed by solving graphically the equation $V' =0$, as illustrated in Fig.~\ref{fig:fig1}. At $\theta = 0$ the minimum is at $\phi=0$ while at $\theta = \pi$ it is at $\phi = \pi$. In both cases $CP$ is unbroken. At $0 < \theta < \pi$ ($\pi < \theta < 2\pi$) the minimum is at some $0 < \phi < \theta$ ($\theta < \phi < 2 \pi$) and $CP$ is explicitly broken.

\begin{figure}
\centerline{\includegraphics[scale=0.45,angle=0]{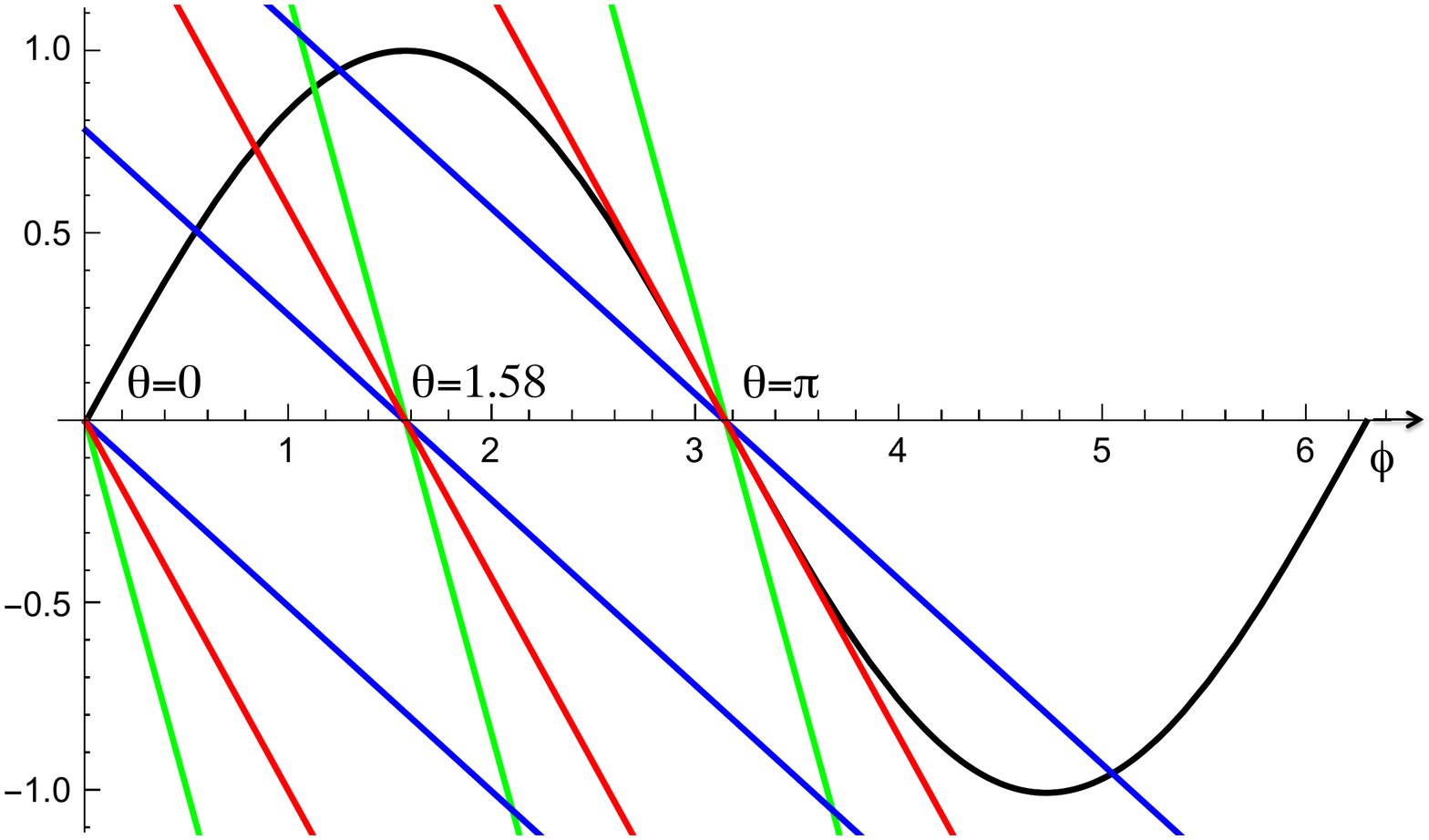}}
\vspace{-1cm}
\caption{\small{Solutions of  $V'(\phi) =0$ are given by the intersections of the curve $\sin \phi$ (black) with the straight lines $(\theta  - \phi)/\epsilon$ for $\theta=0$, $\theta=\pi$ and a generic value taken to be $\theta=1.58$. Code color is as follows: $\epsilon < 1$ green lines, $\epsilon =1$ red lines, $\epsilon > 1$ blue lines.}}
\label{fig:fig1}
\end{figure}

\item{$\epsilon \ge 1$}

This case is much richer. Since now $V''$ can be negative, some stationary points can correspond to maxima rather than minima of $V$.
For a zero mass ground state we should require $V' = V'' = 0$. But for it to be the absolute minimum we should also have $V''' =0$ and $V'''' >0$. However, from~(\ref{potentialder}) we see that $V''' = 0$ is only possible if $\phi = \pi$ mod($\pi$) and therefore (from the first and last of Eqs.~(\ref{potentialder})) if $\theta = \pi$. Let us then consider this case in more detail.

For $\theta = \pi$ there is always a stationary point at $\phi = \pi$ which, however, for the case $\epsilon > 1$, corresponds to a maximum ($V'' <0$). Since $V$ is bounded from below there should be minima elsewhere. Indeed, for $\epsilon = 1 + \delta, \delta \ll1$, one easily finds two (degenerate) minima. For $\epsilon = 1$ the three stationary points degenerate at $\phi = \pi$ and the stable minimum corresponds to a massless $CP$ conserving ground state. 

\begin{figure}
\centerline{\includegraphics[scale=.5,angle=0]{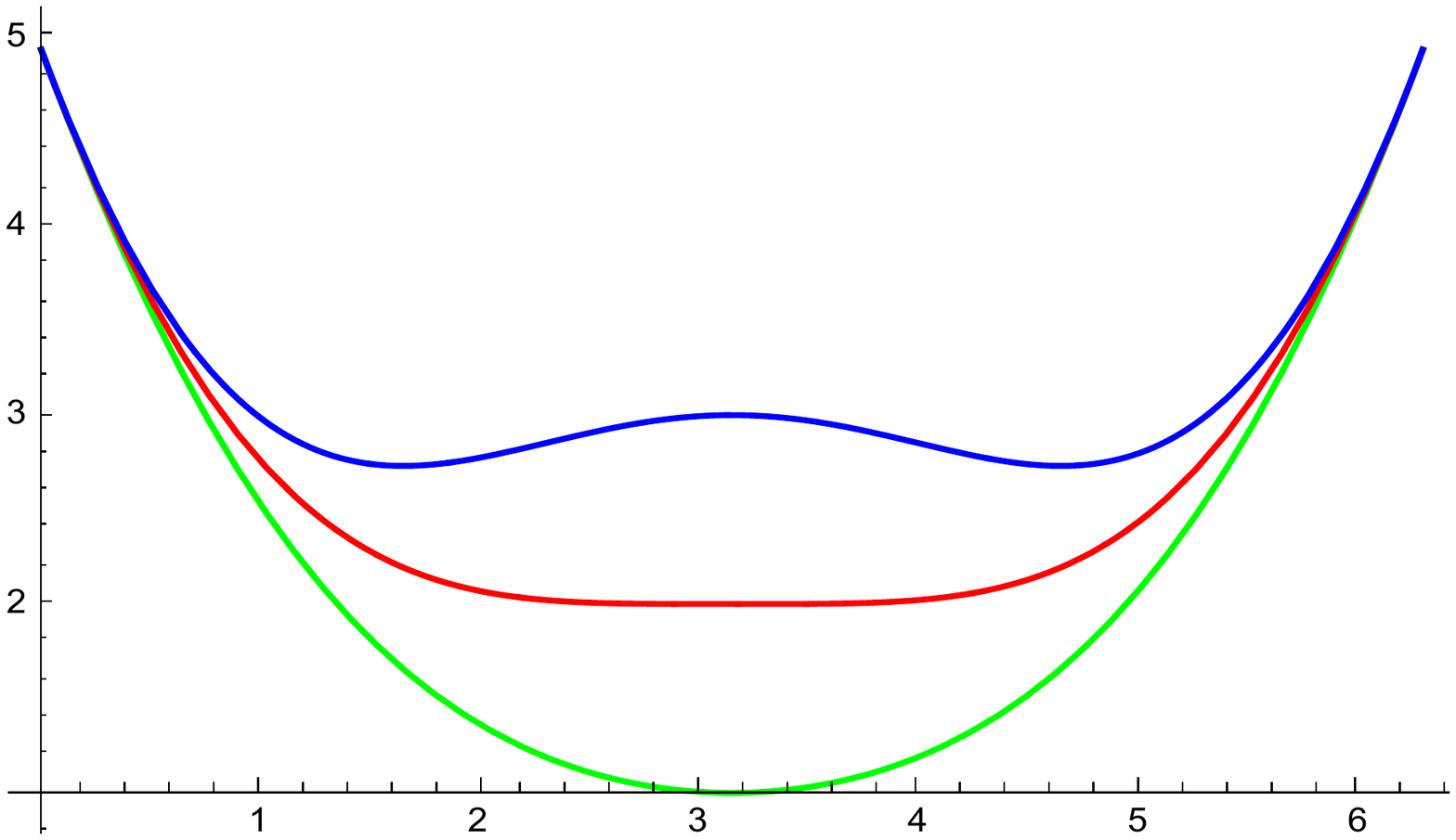}}
\vspace{-1.cm}
\caption{\small{$V(\phi)$ of Eq.~(\ref{potential}) at $\theta=\pi$,  and  $\epsilon=0.5$ (green curve), $\epsilon=1.0$ (red) and  $\epsilon=2.0$ (blue).}}
\label{fig:fig2}
\end{figure}

\begin{figure}
\centerline{\includegraphics[scale=0.55,angle=0]{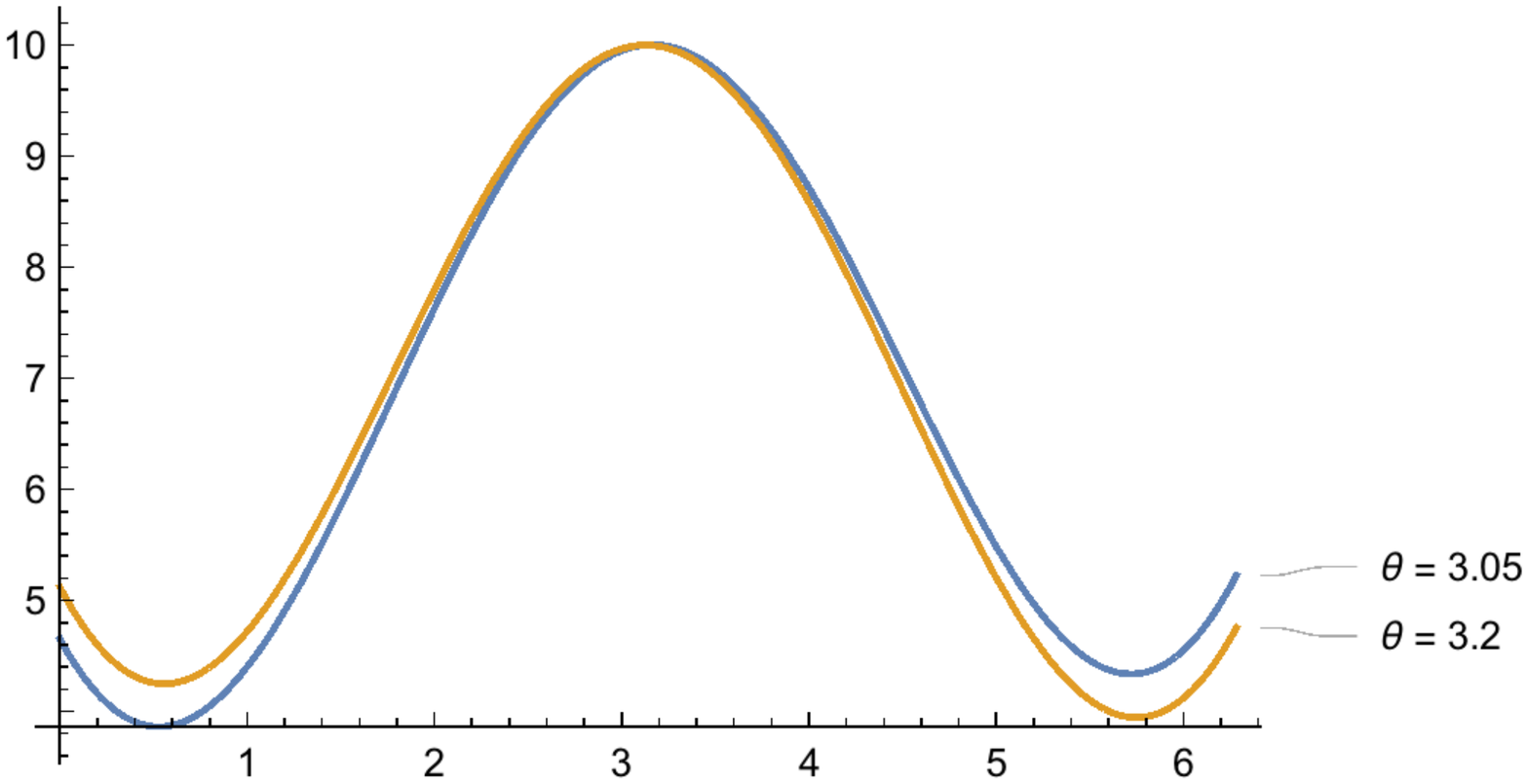}}
\vspace{-3.cm}
\caption{\small{$V(\phi)$ for two values of $\theta$ on opposite sides of $\pi$ and  $\epsilon = 5 $. The true minimum swaps abruptly as one goes through $\theta = \pi$. For the apparent lack of $2\pi$  periodicity in this figure see the discussion in the  text.}}
\label{fig:fig3}
\end{figure}

To make the discussion more quantitative let us assume that $\theta=\pi$ and that $\phi = \pi -\delta$ where
$\delta$ is a small quantity. We can determine $\delta$ by plugging it into the first equation in~(\ref{potentialder}) getting
\begin{eqnarray}
 \delta \left( \frac{\delta^2 \epsilon }{6} +1- \epsilon \right) =0 \, .
\label{deltasm}
\end{eqnarray}
In this way we find again the solution $\delta=0$, which corresponds to a maximum, together with two stable minima related by $CP$ (see below) at 
\begin{eqnarray}
\delta_\pm = \pm \sqrt{ \frac{6 (\epsilon-1)}{\epsilon}} \, .
\label{maxmin}
\end{eqnarray}
This can be seen by plugging~(\ref{maxmin}) in the second of the equations~(\ref{potentialder}) obtaining respectively 
\begin{eqnarray}
\frac{V''}{a}\Big{|}_{\delta = 0} = 1- \epsilon~~~;~~~\frac{V''}{a}\Big{|}_{\delta_{\pm}} = 2(\epsilon -1) \, .
\label{secderi}
\end{eqnarray}
This implies that the solution with $\delta =0$ is a stable one for $\epsilon \leq 1$, while the two other solutions are stable for $\epsilon > 1$ (see Fig.~\ref{fig:fig2}). At $\epsilon=1$ there is a second order phase transition where the PNGB becomes massless. Indeed the mass square is given by the second derivative of the potential computed at the minimum, yielding 
\begin{eqnarray}
M^2 = \mu^2 (\theta) +a = \mu^2 \cos \phi +a \, ,
\label{M2}
\end{eqnarray}
as follows from~(\ref{detAp2=0}) with $N_f=1$. Notice that $M^2$ goes to zero for $\epsilon = 1, \theta = \phi = \pi$.

If we move away from $\theta = \pi$ while $\epsilon >1$ we can have different situations. Below a critical $\epsilon(\theta)$ there is only one minimum while above it an extra couple of stationary points pops out.  One of them is a local maximum, the other a local minimum. Which is the absolute minimum depends on $\theta$. For $\theta < \pi$ the true minimum is at $\phi < \theta$ while for $\theta >  \pi$ it is at $\phi > \theta$ as illustrated in Fig.~\ref{fig:fig3}. Precisely at $\theta = \pi$ there is a two-fold degeneracy easily understood as due to the spontaneous breaking of $CP$~\footnote{Indeed the two minima appear to be symmetric with respect to $\phi = \pi$ and become equal and opposite after a trivial $2 \pi$ shift of one of them.}. This abrupt change in the minimum of the potential around $\theta = \pi$ signals a first order phase transition all along the line
$\mu^2 e^{i \theta} = [- \infty, -a^2]$ ending at the second order phase transition point $\theta = \pi, \mu^2 = a$ as first observed in~\cite{MC2} and more recently discussed in~\cite{Nati,GKS}. 

The second order phase
transition is not only signalled by the mass gap going to zero, but also from 
the divergence of the topological susceptibility (generally defined as the $\langle Q~Q \rangle$ correlator at zero momentum) at $\epsilon=1, \theta = \pi$.
This follows from Eq.~(\ref{chiQCD}) for $N_f =1$
\begin{eqnarray}
\chi_{QCD} = \frac{\chi_{YM}}{1 + \frac{a}{\mu^2 (\theta)} } = \frac{\chi_{YM} \epsilon \cos \phi}{1+ \epsilon \cos \phi} \, ,
\label{chiQCD1}
\end{eqnarray}
which diverges for $\epsilon = 1$  at  $\theta = \phi = \pi$. 

Figs.~\ref{fig:fig2} and~\ref{fig:fig3} illustrate the shape of the potential for different values of $\epsilon$ and for $\theta = \pi$ or $\theta \ne \pi$, respectively. Note that the potentials shown in Figs.~\ref{fig:fig2} and~\ref{fig:fig3} do not look periodic in $\phi$ while they should. Indeed the potential is multi valued because of the log term in the effective Lagrangian~(\ref{ELLE}) and the correct branch has to be chosen as we vary $\phi$. Periodicity is  thus restored at the expense of non-analyticity points (cusps)  in $V$ at particular values of $\phi$. For instance, for $\theta = \pi$ (Fig.~\ref{fig:fig2}) the cusp are at $\phi = 0$ mod($2 \pi$), while for a generic $\theta$  they are at $\theta + \pi$ mod($2 \pi$). 
\end{itemize}

\subsection{$N_f =2$}
\label{Nf=2}

In the case $N_f =2$ with unequal masses (say, $\mu_1^2 < \mu_2^2$) the equations to be solved are
\begin{eqnarray}
\epsilon_1 \sin \phi_1 = \epsilon_2 \sin \phi_2 = \theta - \phi_1 - \phi_2~~~;~~~\epsilon_i \equiv \frac{\mu_i^2}{a}\, .
\label{N21}
\end{eqnarray}
For $\theta=\pi$ the solutions are simply 
\begin{eqnarray}
\phi_1 = \pi ~~;~~~ \phi_2 =0 ~~ {\rm or}~~ \phi_1 = 0  ~~;~~~ \phi_2 = \pi \, .
\label{N22}
\end{eqnarray}
The masses of the two pseudoscalar mesons can be read from Eq.~(\ref{detA}) and are given by
 \begin{eqnarray}
M_{1,2}^2 = a + \frac{\mu_1^2(\theta)  + \mu_2^2 (\theta)}{2} \pm 
\sqrt{ a^2 + \left( \frac{\mu_1^2(\theta) - \mu_2^2 (\theta)}{2} \right)^2} \, ,
\label{FU12}
\end{eqnarray}
valid for arbitrary $\theta$. It is easy to check that the mass squared  with the minus sign 
is massless if the following condition is satisfied
\begin{eqnarray}
a (\mu_2^2 (\theta)  + \mu_1^2 (\theta)) = \left( \frac{\mu_1^2  (\theta)
- \mu_2^2 (\theta)}{2}
\right)^2 -  \left( \frac{\mu_1^2 (\theta) + \mu_2^2 (\theta)}{2}\right)^2 \, .
\label{FU13}
\end{eqnarray}
Notice that, if both $\mu_{1,2}^2 (\theta)$ are positive, the previous condition cannot be satisfied because the r.h.s.\  is always negative, while the l.h.s.\ is always positive. In particular, it cannot be satisfied at $\theta =0$. But at $\theta = \phi_1 = \pi$, the previous condition becomes
\begin{eqnarray}
a (\mu_2^2 - \mu_1^2 ) = \mu_1^2 \mu_2^2    \Longrightarrow
\frac{1}{a} + \frac{1}{\mu_2^2} = \frac{1}{\mu_1^2} \, .
\label{FU14}
\end{eqnarray}
This means that, if the condition 
 \begin{eqnarray}
\frac{1}{\mu_1^2} - \frac{1}{\mu_2^2} \geq \frac{1}{a} 
\label{FU15}
\end{eqnarray}
is fulfilled, $CP$ is unbroken because $\theta - \phi_1 - \phi_2=0$. Although the second solution in~(\ref{N22}) conserves $CP$, it does not correspond to the absolute minimum and does not satisfy~(\ref{FU13}).

On the other hand, if $\mu_1^{-2}  < \mu_2^{-2} + a^{-1}$ not even the first solution in Eq.~(\ref{N22}) corresponds to a minimum and other solutions takes over.  
As in the case $N_f =1$, let us consider the following example. Defining
\be
\epsilon_i = \mu_i^2/a~~;~~ \rho = \epsilon_1/\epsilon_2~~;~~ \sigma = \epsilon_1 +\rho - 1\, ,
\label{Nf2defs}
\ee
one finds, to leading order in $\sigma \ll 1$, the two further solutions 
\be
\phi_1 = \pi - \delta_1~~;~~ \phi_2 = \delta_2~~;~~ \delta_1 = \pm \sqrt{\frac{6 \sigma}{1-\rho^3}}~~;~~ \delta_2 = \rho \delta_1 \,. 
\label{Nf2slns}
\ee

\begin{figure}
\centerline{\includegraphics[scale=0.80,angle=0]{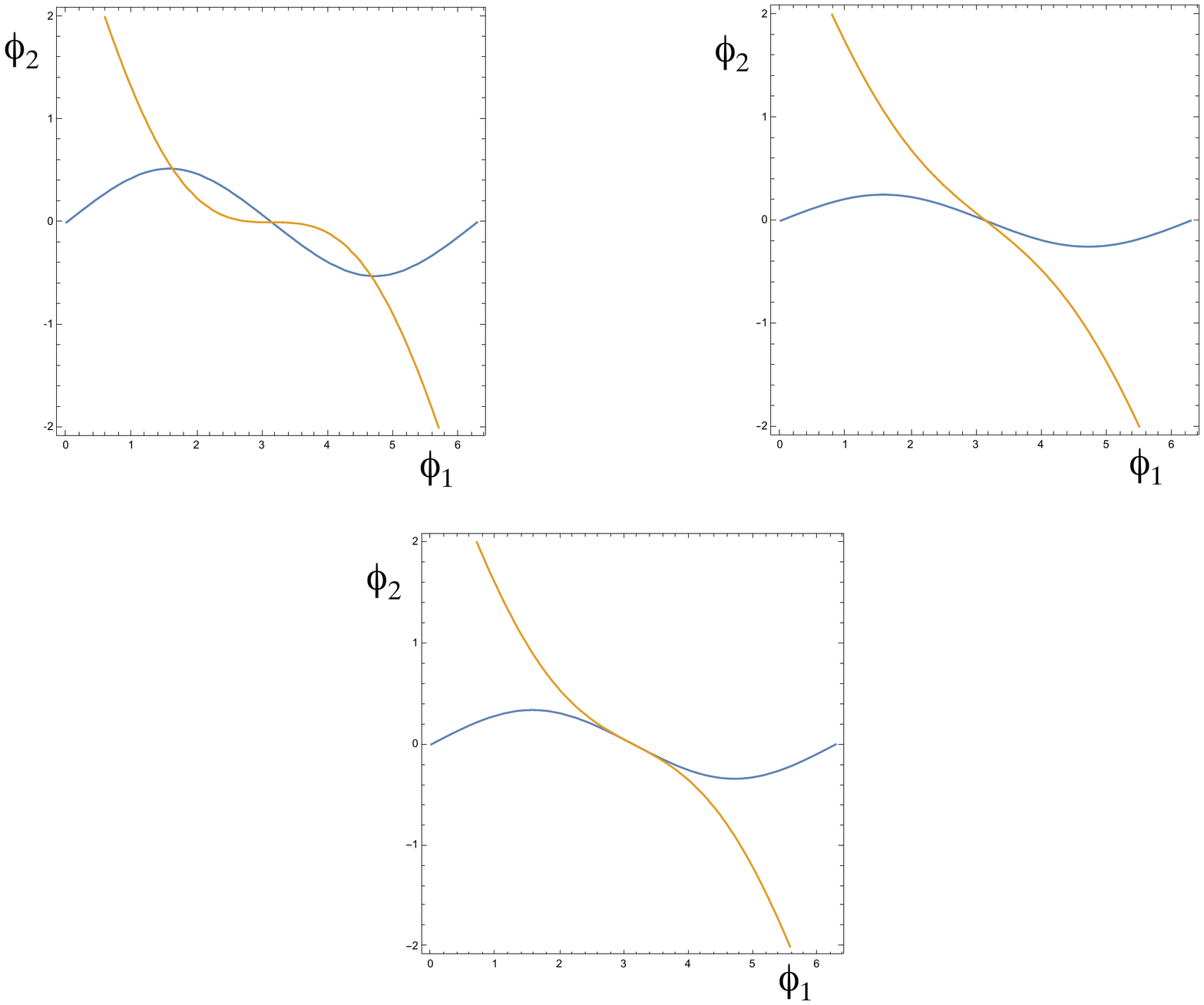}}
\vspace{-.5cm}
\caption{\small{Solutions of the stationarity conditions for $N_f=2$, $\mu_d^2 = 2 \mu_u^2$ and $\theta = \pi$ are given by the intersections of the curves shown in different color. The two situations with one or three solutions are shown together with the limiting case corresponding to a second order phase transition.}}
\label{fig:fig4}
\end{figure}

\begin{figure}
\centerline{\includegraphics[scale=0.70,angle=0]{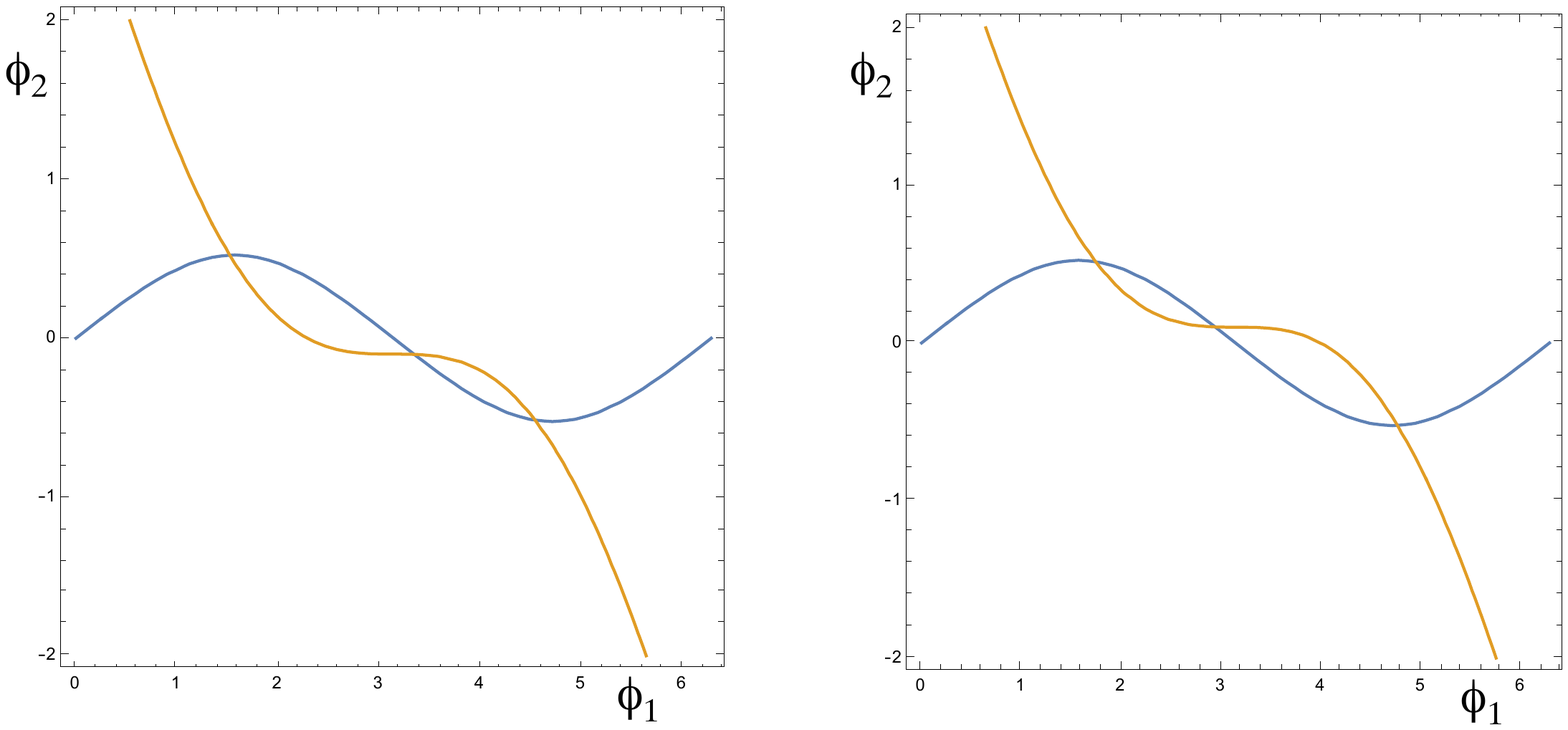}}
\vspace{-3.5cm}
\caption{\small{Same as Fig~\ref{fig:fig4} in the $CP$ broken situation, but for two values of $\theta$ on opposite sides of $\theta = \pi$: (a): $\theta < \pi$, (b): $\theta > \pi$. The true minimum (corresponding to the intersection which is farther away from the middle one) swaps abruptly as one goes through $\theta = \pi$.
}}
\label{fig:fig5}
\end{figure}

In the general case the solutions can be found numerically. Fig.~\ref{fig:fig4} illustrates again the three distinct cases for $\theta=\pi$, while Fig.~\ref{fig:fig5} 
does the same for $\theta \ne \pi$. We see clearly that, as in the $N_f=1$ case, the critical surface $\mu_1^{-2}  = \mu_2^{-2} + a^{-1} $ separates the situation with a single solution from the one with several solutions. In the latter case $CP$ is spontaneously broken and the ground state jumps as we go from $\theta < \pi$ to $\theta >\pi$.
On the critical surface there is a massless excitation and the QCD topological 
susceptibility blows up.

In this generic case the phase structure resembles the $N_f=1$ case. In the complex $\mu_1^2 e^{i\theta}$ plane ($\mu_1^2$ is the smallest mass parameter) we find a line of first order transitions along the negative axis ending on a second order transition point where one mass goes to zero. The position of the second order point depends on the other parameters (mass ratios, $a$). We can also see this structure in the complex $\det \mu^2$ plane, as discussed in the next subsection.

Let us close with a short discussion of the peculiarities of the equal mass case, $\mu_1^2=\mu_2^2=\mu^2$. In this case the condition~(\ref{FU15}) cannot be satisfied except, asymptotically, if we send $\mu^2/a$ to zero. In other words, as discussed in~\cite{GKS}, the first order phase transition line now extends over the whole negative real axis terminating at the origin. However, before jumping too quickly to this conclusion we should observe that the potential becomes very flat for small $\mu^2/a$, so much that it develops a flat direction at O$(\mu^2/a)$. This continuous vacuum degeneracy is lifted at O$((\mu^2/a)^2)$ so that the $CP$ violating minimum is found to lie O$((\mu^2/a)^2)$ below the $CP$ conserving one. The existence of this quasi-flat direction and its lifting to O$(m^2)$ was first pointed out in~\cite{Smilga:1998dh} and further discussed in~\cite{GKS}. In general, O$(m^2)$ corrections are not included in  effective Lagrangians like~(\ref{InitialL3}) but, in the context of our double limit $m/\Lambda \rightarrow 0, N \rightarrow \infty$ with $mN/\Lambda$ fixed (recall $a\sim \Lambda^2/N$), the split in the potential between the two vacua is of order $\Lambda^4(m N/\Lambda)^2$ while the O$(m^2)$ corrections we are ignoring are at least a factor $1/N$ lower. 
We can thus conclude that, above a sufficiently large $N$, $CP$ is broken for two equal mass flavors~\footnote{We thank Z.\ Komargodski for having raised with us the issue of flat directions and for useful correspondence about it.}.

\subsection{$N_f \geq 3$}
\label{Nf}

For a generic number of flavors we have to solve Eqs.~(\ref{minimum}). It can be immediately
seen that for $\theta=\pi$ we have the following solution that generalizes to $N_f$ flavors 
what we found for two flavors, namely~\footnote{We can find many other stationary solution that preserve $CP$ by choosing an arbitrary number of $\phi_i$ to be $\pm \pi$ with their sum adding up to $\theta = \pi$. However, it is trivial to show that the solution in Eq.~(\ref{Nf1}) is, among those, the one with the lowest energy and thus the one to be compared with other (in general $CP$ breaking) solutions.} 
\begin{eqnarray}
\phi_1 = \pi~~;~~~\phi_2 = \phi_3 = \dots = \phi_{N_f} =0~~;~~ \mu_1^2 \le \mu_i^2~~{\rm for}~ i \ne 1 \, .
\label{Nf1}
\end{eqnarray}
It can be immediately checked that the determinant in Eq.~(\ref{detAp2=0})  is positive if the condition  
\begin{eqnarray}
\Delta \equiv \frac{1}{\mu_1^2} - \frac{1}{a} - \sum_{i=2}^{N_f} \frac{1}{\mu_i^2}  > 0 
\label{Nf2}
\end{eqnarray}
is satisfied. In the corresponding region of parameter space 
we have a $CP$ conserving stable solution since
$\theta - \sum_{i=1}^{N_f} \phi_i =0$.  On the surface where~(\ref{Nf2}) is replaced by an equality, the topological susceptibility diverges,
as follows from Eq.~(\ref{chiQCD}), and there is a massless state,
signalling a second order phase transition.
In the region where, instead, $\Delta < 0$, 
the solution in Eq.~(\ref{Nf1}) ceases to be a minimum  and we have to look for new solutions corresponding to minima where we will find that $CP$ is  spontaneously broken. 
 
In terms of the dimensionless quantities 
\be
\epsilon_i \equiv \frac{\mu_i^2}{a}~~,~~ \rho_i  \equiv \mu_1^2/\mu_i^2 = \epsilon_1/\epsilon_i ~~,~~ (i = 2, \dots N_f) ~~;~~  (0 \le  \rho_i \le 1) \, ,
\ee
the criticality condition can be  written as
\be
\epsilon_1 = 1 - \Sigma~~;~~ \Sigma \equiv \sum_{i=2}^{N_f} \rho_i 
\label{critcond}
\ee
and the zero-mass eigenvector  is simply given by 
\be
{\cal V}(M=0) \propto  (1, - \rho_2, \dots, - \rho_{N_f}) \, .
\label{M=0eigenvector}
\ee
Clearly the above expression is consistent with decoupling when one of the $\rho$'s goes to zero. $CP$ is broken (unbroken) when the l.h.s.\ of (\ref{critcond}) is larger (smaller) than the r.h.s. It is always broken if $\Sigma >1$.
If instead we look at the equal mass case, $\rho_i =1$, we see that $\Delta <0$ except in the case $N_f=1$ and $\mu^2 /a < 1$ and in the case $N_f =2$ and $\mu =0$~\cite{GKS}.

As before, in the generic mass case we have a line of first order transition  in the complex $\mu_1^2 e^{i\theta}$ plane ending on a second order point where one physical mass goes to zero. The position of the second order point resides  at the intersection of the negative $\mu_1^2$ line with the critical hyper surface and therefore depends on the other parameters (mass ratios, $a$). 

\begin{figure}[htbp]
\centerline{\includegraphics[scale=0.45,angle=0]{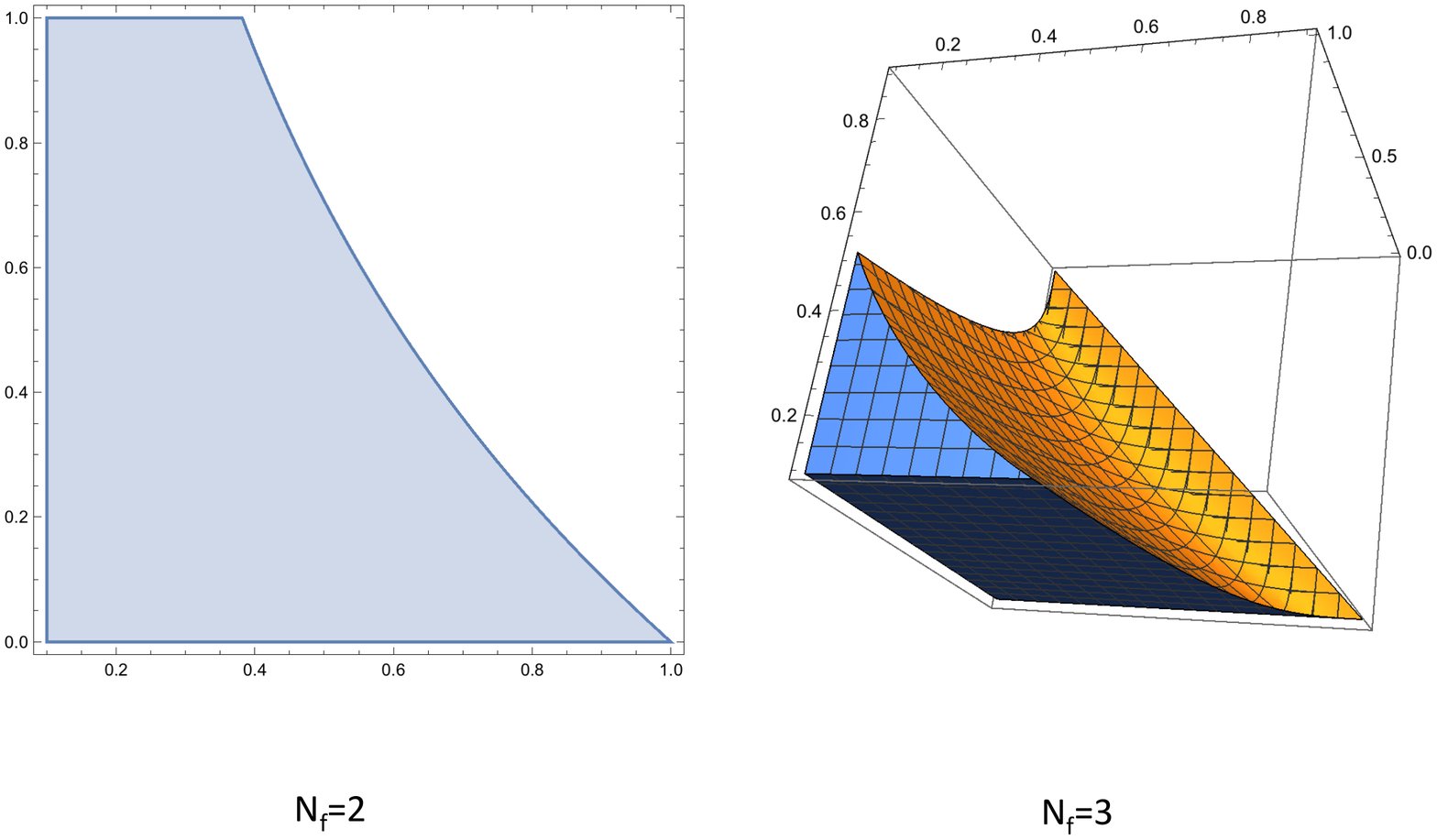}}
\vspace{-1cm}
\caption{\small{$CP$ conserving (filled) and $CP$ breaking (empty) regions for  $N_f = 2$ and $N_f= 3$. The vertical axis is $D$, the horizontal is (are) the mass ratio(s).}}
\label{fig:fig6}
\end{figure}

We end this section giving a definition of the critical hypersurface in terms of the quantity $D \equiv \det(\mu^2/a^2) = \det(\epsilon)$, where, however,  $\mu^2$ is now the matrix introduced in~(\ref{InitialL3}) {\it after} having absorbed the $\theta$ angle by a chiral rotation~\footnote{We recall that only $\bar{\theta} = \theta + \arg D$ is physically relevant. In the rest of the paper we adopted the convention of having $\mu^2$ diagonal, real and positive and $\theta$ arbitrary. For the rest of this section, instead, $\theta =0$ and $\arg D$ is arbitrary.}. 
The critical value of $D$, $D_{c}$, is negative (corresponding to $\bar{\theta} = - \arg D = \pm\pi$) and its absolute value depends only on the ratios $\rho_i$ introduced earlier.
 Indeed the condition for $CP$ violation can be expressed as follows
\be
|D| >  |D_c|~~;~~  |D_c^{1/N_f}| = (1 - \Sigma) \Pi^{-1/Nf}  ~~;~~ \Pi = \prod_{i=2}^{N_f} \rho_i \, .
\ee     
It can be checked that also the above expression satisfies decoupling when  one of the $\rho_i$'s goes to zero.
Plots of the $CP$ conserving regions and of the critical lines (surfaces) for $N_f = 2$ ($N_f= 3$) are shown in Figs.~\ref{fig:fig6}.

\section{Spontaneous $CP$ violation and the axion potential}
\label{axion}
\setcounter{equation}{0}

We shall now discuss some consequences of the considerations made in the previous sections when an extra dynamical low-energy degree of freedom, the axion, is added to those of chiral QCD. As pointed out independently by Weinberg~\cite{SWax} and Wilczek~\cite{FW}, 
the existence of an axion is a necessary consequence of the Peccei-Quinn (PQ) resolution~\cite{PQ} of the strong-$CP$ problem. The latter consists in the observation that present bounds on the electric dipole moment of the neutron force the $\theta$ angle (actually $\bar{\theta}$) to be less than $10^{-9}$~\cite{Crewther:1979pi}.
 
Of course, if one of the quarks is massless, the strong-$CP$ problem would be automatically solved since $\theta$ could be rotated away (equivalently $\bar{\theta} =0$). Unfortunately, the low-energy spectrum of QCD is inconsistent with the data if one of the quark flavors is massless. A generic way to introduce the PQ resolution of the problem, and the axion, parallels the massless quark solution while avoiding its unwanted consequences. One assumes the existence a new axial 
$U(1)$ global symmetry, only broken by the QCD anomaly (in QCD that symmetry would be the chiral rotation of the massless quark field). Then the existence of the axion follows from Goldstone's theorem associated with the spontaneous breaking of this symmetry. The axion is only a PNGB because there is an no anomaly-free spontaneously broken exact symmetry. The only additional free-parameters with respect to QCD are the so-called axion decay constant $F_{\alpha}$, the analog of $F_\pi$, and $\alpha_{PQ}$, denoting the strength of the contribution of the new sector to the $U_A(1)$ anomaly. Instead, the $\theta$ parameter can be rotated away as we shall now discuss in detail.

\subsection{Including the axion in the QCD effective Lagrangian}

In view  of the above considerations, the axion can be easily incorporated in the QCD
effective Lagrangian discussed in Sect.~\ref{effective} as if there were an extra zero-mass fermion, condensing at the scale $F_{\alpha}$, and contributing to the anomaly with a coefficient $\alpha_{PQ}$ (relative to the weight of a QCD fermion). This can be simply  implemented by introducing, together with $U$ and $\Phi$,  
similarly related axionic fields $\alpha$ and $N$
\begin{eqnarray}
N = \frac{F_\alpha}{\sqrt{2}}
{\rm e}^{i \frac{\sqrt{2}}{F_{\alpha}} \alpha} \, .
\label{UN2b}
\end{eqnarray}
The generalization of the Lagrangian~(\ref{InitialL3}) then reads~\footnote{See Ref.~\cite{DVS}.}
\begin{eqnarray}
&&L = \frac{1}{2} {\rm Tr} \left( \partial_\mu U \partial^\mu U^{\dagger} 
\right) + \frac{1}{2}  \partial_\mu N \partial^\mu N^{\dagger} +
 \frac{F_\pi}{2\sqrt{2}} {\rm Tr} \left( \mu^2 (U+U^\dagger) \right)  + 
\frac{Q^2}{2 \chi_{YM}}  
 \nonumber \\
&&  + \frac{i}{2} Q \left[ \log U - \log U^\dagger + \alpha_{PQ} \left( \log N - 
\log N^\dagger \right)
\right] - \theta Q\, .
\label{InitialL3b}
\end{eqnarray}

Restricting, for the sake of simplicity, our analysis to the fields in the Cartan sub-algebra
of the QCD pseudoscalar mesons, the previous Lagrangian becomes 
\begin{eqnarray}
&&L = \frac{1}{2} \sum_{i=1}^{N_f} \partial_\mu v_i \partial^\mu v_i  + 
\frac{F_\pi^2}{2} \sum_{i=1}^{N_f} \mu_i^2 \cos \left(  -\phi_i + \frac{\sqrt{2}}{F_\pi} v_i
\right)
+ \frac{Q^2}{2 \chi_{YM}} \nonumber \\
&&  +  \frac{1}{2} (\partial_\mu \alpha )^2 - Q \left( \theta - \sum_{i=1}^{N_f} \phi_i - \beta +
\frac{\sqrt{2}}{F_\pi} 
 \sum_{i=1}^{N_f}  v_i + \frac{\alpha_{PQ} 
\sqrt{2}}{F_{\alpha}} \sigma   \right) \, ,
\label{L2vvv}
\end{eqnarray} 
where again we have allowed for a non-trivial  expectation $\langle U \rangle$  as in Eq.~(\ref{vi}) and we have also introduced an expectation value for $\alpha(x)$ and a shifted axion field $\sigma$ as $\alpha (x)= - \frac{\alpha_{PQ} \sqrt{2}}{F_{\alpha}} \beta + \sigma (x)$.

Proceeding now as in Sect.~\ref{effective}, we determine the phases $\phi_i$ and $\beta$ by minimizing
\begin{eqnarray}
V (\phi_i, \beta) = - \frac{F_\pi^2}{2} \sum_{i=1}^{N_f} \mu_i^2 \cos \phi_i +
\frac{\chi_{YM}}{2} \left(\theta - \sum_{i=1}^{N_f} \phi_i -\beta \right)^2 \, .
\label{Vphiiphiax}
\end{eqnarray}
The stationary points of this potential are solutions of the equations
\begin{eqnarray}
&&- \frac{F_\pi^2}{2}\mu_i^2 \sin \phi_i + \chi_{YM} 
(\theta - \sum_i \phi_i - \beta)=0 ~~~;~~~i=1, 2, \dots, N_f  \nonumber \\
&& \theta - \sum_i \phi_i - \beta=0 \, ,
\label{minimumax}
\end{eqnarray}
and are given by 
\begin{eqnarray}
\hat\phi_i = 0  \,\,\, \mod \!(\pi)~~;~~~\hat\beta= \theta - \sum_{i=1}^{N_f} \phi_i \, .
\label{moregeneral}
\end{eqnarray}
We notice that the choice 
\begin{eqnarray}
\hat\phi_i =0 ~~;~~i=1, 2, \dots, N_f ~~~;~~~ \hat\beta=\theta 
\label{solu}
\end{eqnarray}
corresponds to the minimum of the potential, while the other choices correspond to maxima or to saddle points.
Setting the expectation values to~(\ref{moregeneral}), Eq.~(\ref{L2vvv}) takes the form
\begin{eqnarray}
&&L = - V(\hat\phi_i , \hat\beta) +\frac{1}{2} \sum_{i=1}^{N_f} \partial_\mu v_i \partial^\mu v_i  + 
\frac{F_\pi^2}{2} \sum_{i=1}^{N_f} \mu_i^2 \cos \left( \frac{\sqrt{2}}{F_\pi} v_i
\right)
+ \frac{Q^2}{2 \chi_{YM}} \nonumber \\
&&  +  \frac{1}{2} (\partial_\mu \sigma )^2 - Q \left( 
\frac{\sqrt{2}}{F_\pi} 
 \sum_{i=1}^{N_f}  v_i + \frac{\alpha_{PQ} 
\sqrt{2}}{F_{\alpha}} \sigma   \right) \, ,
\label{L2vvvw}
\end{eqnarray}
where $V(\hat\phi_i , \hat\beta)$ is a constant. Thus unlike the QCD case, physics has become $\theta$-independent and $CP$ conserving.
As we shall see in the following subsection, the full richness of the QCD case reappears once we consider the axion potential.

For the moment, in analogy with Eq.~(\ref{ELLE1}), we rewrite~(\ref{L2vvvw}) in the form
\begin{eqnarray}
&& L = - V(\hat\phi_i , \hat\beta) + \frac{1}{2} \sum_{i=1}^{N_f} \partial_\mu v_i \partial^\mu v_i  + \frac{F_\pi^2}{2} \sum_{i=1}^{N_f} \mu_i^2 \left(\cos \left( \frac{\sqrt{2}}{F_\pi} v_i
\right) -1 \right)  +  \frac{1}{2} (\partial_\mu \sigma )^2  \nonumber \\
&&- \frac{\chi_{YM}}{2} 
\left( \frac{\sqrt{2}}{F_\pi} \sum_{i=1}^{N_f} v_i + \frac{\alpha_{PQ} \sqrt{2}}{F_\alpha} \sigma
\right)^2 \nonumber \\
&& + \frac{1}{2 \chi_{YM}} \left( Q - \chi_{YM} \left( \frac{\sqrt{2}}{F_\pi} \sum_{i=1}^{N_f} v_i + \frac{\alpha_{PQ} \sqrt{2}}{F_\alpha} \sigma  \right)   \right)^2 \, .
\label{ELLEaxion}
\end{eqnarray}
The mass spectrum of the system can be found by diagonalizing the quadratic part of Eq.~(\ref{ELLEaxion}) which reads
\begin{eqnarray}
\hspace{-.6cm}&& L_2 = \frac{1}{2} \sum_{i=1}^{N_f} \partial_\mu v_i \partial^\mu v_i  - 
\frac{1}{2}\sum_{i=1}^{N_f} \mu_i^2 v_i^2 - \frac{\chi_{YM}}{2} \left(
\frac{\sqrt{2}}{F_\pi} \sum_{i=1}^{N_f} v_i + \frac{\sqrt{2} \alpha_{PQ}
}{F_\alpha} \sigma \right)^2\! \!+\! \frac{1}{2} (\partial_\mu \sigma )^2=
\nonumber \\
\hspace{-.6cm}&&  =  \frac{1}{2} 
\sum_{a=1}^{N_f+1} \partial_\mu H_a \partial^\mu H_a    - \frac{1}{2} H^T A H \, ,
\label{L2v4}
\end{eqnarray}
where $H$ is an $N_f +1$-column vector  and $A$ is the squared-mass matrix 
\begin{eqnarray}
H = \left( \begin{array}{c} \sigma \\ v_1 \\ v_2 \\ \cdot \\ \cdot  \\ v_{N_f} \\ 
\end{array} \right) ~~;~~
A = \left( \begin{array}{cccccc}   b^2 a &  ba &  ba &  ba &  \dots &
 b a \\
     ba  &   \mu_1^2 +a &  a  &  a & \dots &    a \\
   ba  & a &   \mu_2^2 + a & a & \dots &   a \\
\dots &  \dots & \dots  & \dots & \dots & \dots\\
ba & a & a & a &   \dots &   \mu_{N_f}^2 + a  
\end{array} \right)  \, .
\label{VandA}
\end{eqnarray}
The mass spectrum is the result of the diagonalization of $A$ and can be read off from 
\begin{eqnarray}
&&\det \left( p^2 \delta_{ij} - A_{ij} \right) = p^2 \prod_{i=1}^{N_f} 
(p^2 - \mu_i^2) \left[ 1 - a \left( \sum_{i=1}^{N_f} \frac{1}{
p^2 - \mu_i^2} + \frac{b^2}{p^2} \right) \right] \nonumber \\
&&=\prod_{i=1}^{N_f+1}\left( p^2 - M_i^2 \right)\, ,
\label{detAaxion}
\end{eqnarray}
where $a=\frac{2\chi_{YM}}{F_\pi^2}$ (as in Eq.~(\ref{defa})) and $ b = \frac{F_\pi \alpha_{PQ}}{F_{\alpha}}$. The
$M_i$ are the masses of the physical states that diagonalize the mass matrix.
By going to $p^2$ = 0,  Eq.~(\ref{detAaxion}) implies 
\be 
\label{detMaxion} 
\det A = a b^2 \prod_{i=1}^{N_f} \mu_i^2  = \prod_{j=1}^{N_f+1} M^2_j \, ,
\ee 
where the product on the r.h.s.\ includes the axion as well as the Cartan PNGB masses.
Note that, unlike the non-axionic case, for non-vanishing $m_i$, $a$ and $b$, this determinant is always positive implying no massless state (and indeed a non-tachyonic spectrum). This would have also been the case had we considered QCD with one massless flavor (in that case $b=1$). In particular, for small $b$, the mass of
the axion is given by looking for a zero at small $p^2$ of the term in square brackets in 
Eq.~(\ref{detAaxion}). Neglecting $p^2$ with respect to $\mu_i^2$ one obtains
\begin{eqnarray}
M_{axion}^2 = \frac{b^2}{\frac{1}{a}+  \sum_{i=1}^{N_f} \frac{1}{\mu_i^2} } \, .
\label{Maxion}
\end{eqnarray}
This reduces to the usual expression for the axion mass~\cite{SWax,BT} 
 in the limit $a, \mu_s^2 \gg \mu_{u,d}^2$.
Alternatively, using Eq.~(\ref{chiQCD}) and the definition of $b$, we can write 
\begin{eqnarray}
M_{axion}^2 
= \frac{ 2 \alpha_{PQ}^2}{F_{\alpha}^2} \chi_{QCD}\, ,
\label{Maxion1}
\end{eqnarray}
another formula often used in the literature (see e.g.\ Ref.~\cite{Preskill:1982cy}).

Finally, from the term in the last line of Eq.~(\ref{ELLEaxion}) and the matrix definition in Eq.~(\ref{L2v4}) we get (having $\langle Q \rangle$ = 0)  the following two-point correlation function
\begin{eqnarray}
\langle Q(x) Q(y) \rangle^{F.T.} = i \chi_{YM} \frac{p^2 \prod_{i=1}^{N_f} (p^2 - \mu_i^2)}{
\prod_{i=1}^{N_f +1} (p^2 - M_i^2)}= \frac{i \chi_{YM}}{
\left[ 1 - a \left( \sum_{i=1}^{N_f} \frac{1}{
p^2 - \mu_i^2} + \frac{b^2}{p^2} \right) \right] } \, ,
\label{QQaxion}
\end{eqnarray}
that vanishes at $p^2=0$ signalling that the topological susceptibility in a theory 
where QCD is ``augmented'' by another sector that includes the axion, is zero consistently with the fact that the dependence on the $\theta$ parameter disappears.

For the physically interesting case we have to take $b \ll 1$ so that the spectrum should contain a very light pseudo-scalar, the physical axion, which is the original field $\sigma$ 
up to an O$(b)$ admixture of PNGBs. This is all well known.
We will now discuss how things take an interesting turn when we go from properties of the spectrum (i.e.\ of small fluctuations around the minimum of $V$) to those of the full potential at a finite distance from its minimum.

\subsection{The axion potential}

From Eq.~(\ref{ELLEaxion}) we can immediately read the axion-PNGB potential
\begin{eqnarray}
\hspace{-1.2cm}&&V( v_i, \sigma)= - \frac{F_\pi^2}{2} \sum_{i=1}^{N_f} \mu_i^2 
\cos \left( \frac{\sqrt{2}}{F_\pi} v_i\right) +\frac{a}{2}
\left(\sum_{i=1}^{N_f} v_i + b \sigma  \right)^2\, . 
\label{POTENTIAL1}
\end{eqnarray} 
In the literature one introduces the concept of an axion potential after integrating out the remaining $N_f$ degrees of freedom in the assumption that they are much heavier then the axion. In principle this requires diagonalizing the mass matrix so as to be in position of identifying the lowest lying state, the physical axion that will be a mixture of $\sigma$ and the $v_i$. In the limit of very small $b$, which is where physics lies, one can neglect these mixings and identify $\sigma$ with the axion modulo some exceptional cases to be discussed below.

For the physically interesting case of two light flavors the axion potential was first derived in~\cite{DVV} under the assumption $\mu_1^2, \mu_2^2 \ll a$ with the result~\cite{diCortona:2015ldu} 
\begin{equation}
V_{axion}(\sigma)=-  \frac{F_\pi^2}{2}\sqrt{(\mu^2_1+\mu^2_2)^2-4\mu^2_1\mu^2_2\sin^2\left(\frac{\alpha_{PQ} \sigma}{\sqrt{2} F_{\alpha}}\right)}+{\mbox{O}}(\mu_i^2/a)\, ,
\label{MINV5}
\end{equation}
which for $N_f =1$ simply becomes
\begin{equation}
V_{axion}(\sigma)=-  \frac{F_\pi^2}{2}~\mu^2  \cos \left( \frac{\sqrt{2}  \alpha_{PQ} \sigma}{F_{\alpha}}\right)+{\mbox{O}}(\mu^2/a)\, .
\label{VNf1}
\end{equation}
We see, however, that by having considered the axion potential at a generic value of  $\sigma$ we have effectively recovered, {\it mutatis mutandis}, the situation discussed in QCD at fixed $\theta$. This is why the discussion of Sect.~\ref{QCDPD} becomes very relevant here. Indeed, the previous analysis shows that, precisely around $\sigma = \frac{\pi F_{\alpha} }{\sqrt{2} \alpha_{PQ}}$, some PNGB mass can become arbitrarily small. In this case integrating out the PNGB fields is no longer justified and a more careful analysis is needed. In other cases the naive solution for the $v_i$ corresponds to a maximum and it has to be replaced with the right solution. The rest of this section is devoted to such an analysis for different numbers of quark flavors. 

In the following, for simplicity of notation, we shall denote by $\varphi_i$ and $\zeta$ the dimensionless quantities $-\frac{\sqrt{2}}{F_\pi} v_i$ and $\frac{\sqrt{2} 
 \alpha_{PQ}}{F_{\alpha}}\sigma$, respectively. In this notation the potential~(\ref{POTENTIAL1}) simply reads
\begin{eqnarray}
\hspace{-1.2cm}&&  2 F_{\pi}^{-2} V(\zeta, \varphi_i) = - 
 \sum_{i=1}^{N_f} \mu_i^2 
\cos \varphi_i +\frac{a}{2}
\left(\sum_{i=1}^{N_f} \varphi_i - \zeta  \right)^2\, . 
\label{POTENTIAL1bis}
\end{eqnarray} 

\subsubsection{$N_f = 1$}
\label{Nf=1axion}

The potential ${V}(\zeta, \varphi)$ has two distinct stationary points, one at $\zeta = \varphi = 0$ and one at $\zeta= \varphi = \pi$. The first is a true minimum, the second a saddle point.
Let us now consider the stationary points in $\varphi$ at fixed $\zeta$ in order to compute $V_{axion}(\zeta)$, distinguishing three cases (looking at Fig.~\ref{fig:fig1} can help following the discussion).

\begin{itemize}
\item $\mu^2/a <1$. In this case there is a single stationary point at $\hat{\varphi}(\zeta) \le \zeta$  which grows monotonically with $\zeta$ interpolating between the two stationary points of $V$. In this case the potential~(\ref{VNf1}) is easily recovered. At $\zeta = \pi$ the potential is smooth and reaches a maximum lying $\mu^2 F_{\pi}^2$ above the absolute minimum. One can easily check that, for $\mu^2/a$ not too close to 1, the mass of the PNGB is always much larger than the scale of variation of the axion potential so that integrating out that degree of freedom is justified. We shall discuss separately the case $|1-\mu^2/a| \ll1$.

\item $\mu^2/a >1$. In this case, as one varies $\zeta$ from $0$ to $\pi$, $\hat{\varphi}(\zeta)$ remains always smaller than $\zeta$. Actually, above a value of $\zeta$ that depends on $\mu^2/a$, new stationary points in $\varphi$ (lying above $\varphi = \pi$) appear but they have higher energy. This is nothing but the situation we have described and discussed around Fig.~\ref{fig:fig3}. In particular, as we approach $\zeta = \pi$, $\hat{\varphi}$ approaches a finite value smaller than $\pi$ and behaving as  $\pi a/\mu^2$ for $\mu^2/a \gg 1$. Precisely at $\zeta = \pi$ this minimum becomes degenerate with one at $\varphi >\pi$ which, upon a shift by $2 \pi$ is just its $CP$ transformed.
Again, for  $\mu^2/a$ not too close to 1, integrating out the PNGB appears fully justified but, instead of~(\ref{VNf1}), we get 
\begin{equation}
V_{axion}(\sigma)= \frac12 \chi_{YM}  \left( \frac{\sqrt{2}  \alpha_{PQ} \sigma}{F_{\alpha}}\right)^2 + {\mbox{O}}(a/\mu^2)\, ,
\label{VNf1bis}
\end{equation}
where for a moment we have reintroduced the canonical $\sigma$ field. In particular, the axion mass is now controlled by $a$ rather than by $\mu^2$. At the boundary of its periodicity interval $V_{axion}$ now reaches its maximal value $ \frac12 \chi_{YM} \pi^2 \ll \mu^2 F_{\pi}^2$ (in the small-$a$ limit). Furthermore, at that point its first derivative is non-vanishing (and positive) and, since the potential is periodic, its first derivative will be discontinuous, giving a spike at $\zeta = \pi$. This, of course, is related to the fact that the solution for $\hat{\varphi}$ jumps abruptly as we go through $\theta = \pi$ (see again Fig.~\ref{fig:fig3}).

\item $|1-\mu^2/a| \ll 1$. This third regime is perhaps the most interesting one, at least theoretically. Let us consider the mass matrix (better the matrix of second derivatives) around $\zeta = \varphi = \pi$. It takes the form
\begin{eqnarray}
A  = \left( \begin{array}{cc}   b^2 a &  ba \\
     ba  &  - \mu^2 +a  
\end{array} \right)  \, .
\label{ANf=1}
\end{eqnarray}

We see that, if $|\mu^2 - a| = {\mbox{O}}(ba)$, the off-diagonal entries become of the same order as the difference between the two diagonal ones (remember that $b \ll1$).
This is precisely the situation in which the two eigenvectors are strongly mixed w.r.t. the original (axion-PNGB) basis. Indeed the maximal mixing occurs at $\mu^2 = a(1 - b^2)$ since then the matrix $A$ becomes
\begin{eqnarray}
A  =  \left( \begin{array}{cc}   b^2 a &  ba \\
     ba  &  b^2 a   
\end{array} \right)  \, ,
\label{ANf=1maxmix}
\end{eqnarray}
whose eigenvectors are $(1, \pm1)$,  with eigenvalues $b^2 a \pm ba$. In fact, as we go through the point $\mu^2 =a$,  the two eigenvectors evolve very quickly (i.e.\ as $\mu^2$ goes from $a - {\mbox{O}}(ab)$ to $a +{\mbox{O}}(ab)$) from almost pure axion to almost pure PNGB or vice versa. This is clearly shown by the numerical calculation presented in 
Fig.~\ref{fig:fig7}. Since $\det A < 0$ the spectrum always consists of a normal and a tachyonic state, but the latter is mainly in the PNGB direction at large  $\mu^2$ while it becomes mainly axion-like at small $\mu^2$. That means that, had we started the evolution of the PNGB plus axion system at $\zeta = \varphi = \pi$ the evolution would go immediately towards
smaller $\zeta$'s if $\mu^2 <a$ while, for $\mu^2 > a$, it would first roll down to the true minimum in $\varphi$ and only then will roll down towards $\zeta =0, \varphi =0$.

It is also quite clear that in this particular range of $\mu^2/a$ and $\zeta$ it is not possible to describe the system only in terms of a $V_{axion}(\zeta)$ since the other degree of freedom is as light as the axion itself. Only a description in terms of a $V(\zeta, \varphi)$ is fully adequate.

\end{itemize}

\subsubsection{$N_f \ge 2$ and discussion}

The real world has two very light quarks, $u$ and $d$, a light one, $s$, and three heavy quarks. The latter play no role in our discussion. Thus the case of  physical interest is $N_f =2$ or $3$. Also, at zero temperature, the quantitative solution of the $U(1)$ problem requires~\cite{EW1}, \cite{GV} $\mu_u^2 < \mu_d^2 << \mu_s^2 <  a$. The ratios $\mu_u^2 : \mu_d^2 : \mu_s^2 : a$ are about $1 : 2 : 40 :18$. In what follows we shall  use these numbers together with the results we obtained from the large-$N$ effective action approach, even though in the real world $N=3$. The success of the large-$N$ solution to the $U(1)$ problem suggests that, at least in this sector, the large-$N$ expansion converges quite fast.

We should keep in mind, however, that, while quark mass ratios are expected to be constant below the QCD deconfining temperature (they depend on phenomena occurring at the electroweak-breaking scale), the temperature dependence of $\chi_{YM}$ could possibly differ from that of the quark condensate meaning a possible (strong?) $T$-dependence of $\mu^2/a$.  An increase of that ratio by an order of magnitude would bring us inside the $CP$ broken region.  The available lattice measurements~\cite{Berkowitz:2015aua,Borsanyi:2015cka,Bonati:2013tt}  do not seem to favor this possibility. We defer further comments on this issue to the conclusion section.

In the following we will consider therefore the case of two or three quark flavors of different masses and allow for arbitrary ratios $\mu_i^2/a$.
The situation is now more involved than in the $N_f=1$ case, but qualitatively similar. The stationary points of the potential~(\ref{POTENTIAL1bis}) are 
 \be
 \zeta= 0, \pi \,\,\, {\mbox{mod}}\, (2 \pi)~~;~~ \varphi_i = 0, \pi \,\,\, {\mbox{mod}}\, (2 \pi)~~;~~ \sum \varphi_i = \zeta \, .
 \label{aNfstps}
 \ee
The absolute minimum is as usual the trivial one $\zeta = \varphi_i =0$. 
In general it is legitimate to integrate out the PNGB degrees of freedom by minimizing their potential at fixed $\zeta$ and then insert the solution $\hat{\varphi_i}(\zeta)$ in $V(\zeta, \varphi_i)$. If $\mu_i^2 \ll a$ this can be easily done. In the two-flavor case this gives the result~(\ref{MINV5}).
In the three-flavor case recalling that 
\be
\sin \phi_s = \mu_u^2/\mu_s^2 \sin \phi_u \ll \sin \phi_u \, ,
\ee
we see that the result~(\ref{MINV5}) still holds up to corrections O$(\mu_{u,d}^2/\mu_s^2)$. This is indeed the result used in the literature.

What happens if, for some physical reason, $\chi_{YM}$ drops so fast with $T$ that $a$ becomes of order $\mu_{u,d}^2$ or even smaller? We can understand the situation by considering what happens at the saddle point corresponding to
\be
\zeta = \varphi_u = \pi~,~ \varphi_d = \varphi_s = 0 \, .
\label{ausaddle}
\ee

We have seen in Sects.~\ref{Nf=2} and~\ref{Nf} that the condition for having a massless boson (in the absence of the axion) is
\be
 \frac{1}{\mu_u^2} = \frac{1}{a} +  \frac{1}{\mu_d^2}  + \frac{1}{\mu_s^2} \sim \frac{1}{a} +  \frac{1}{\mu_d^2}~~ \Rightarrow a (\mu_d^2 - \mu_u^2) = \mu_u^2 \mu_d^2 \, .
 \label{cond}
\ee
Precisely around this point we expect a large mixing to occur between the would-be massless  PNGB and the axion and, as one goes through that region, we expect the tachyonic boson to change its dominant component from axionic to mesonic.

\begin{figure}
\centerline{\includegraphics[scale=0.65,angle=0]{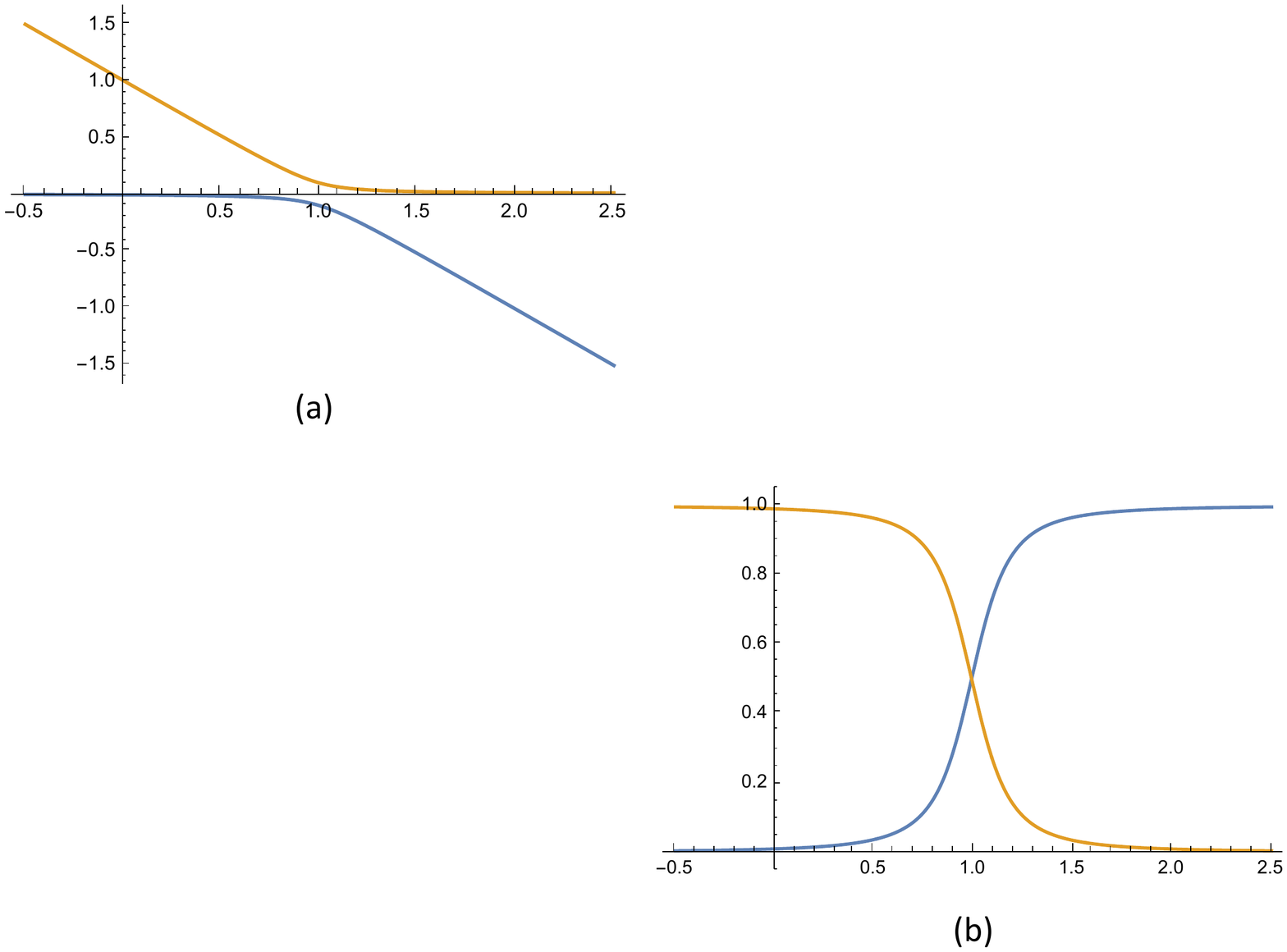}}
\vspace{-.5cm}
\caption{\small{$N_f=1$. (a) Evolution of the two eigenvalues of~(\ref{ANf=1}) for $b = 0.1$ as one varies $\mu^2/a$. The lower eigenvalue is tachyonic. (b) Projections of the two corresponding eigenvectors along the PNGB direction. Maximal mixing occurs in the vicinity of the critical point $\mu^2/a =1$.}}
\label{fig:fig7}
\end{figure}

\begin{figure}
\centerline{\includegraphics[scale=0.65,angle=0]{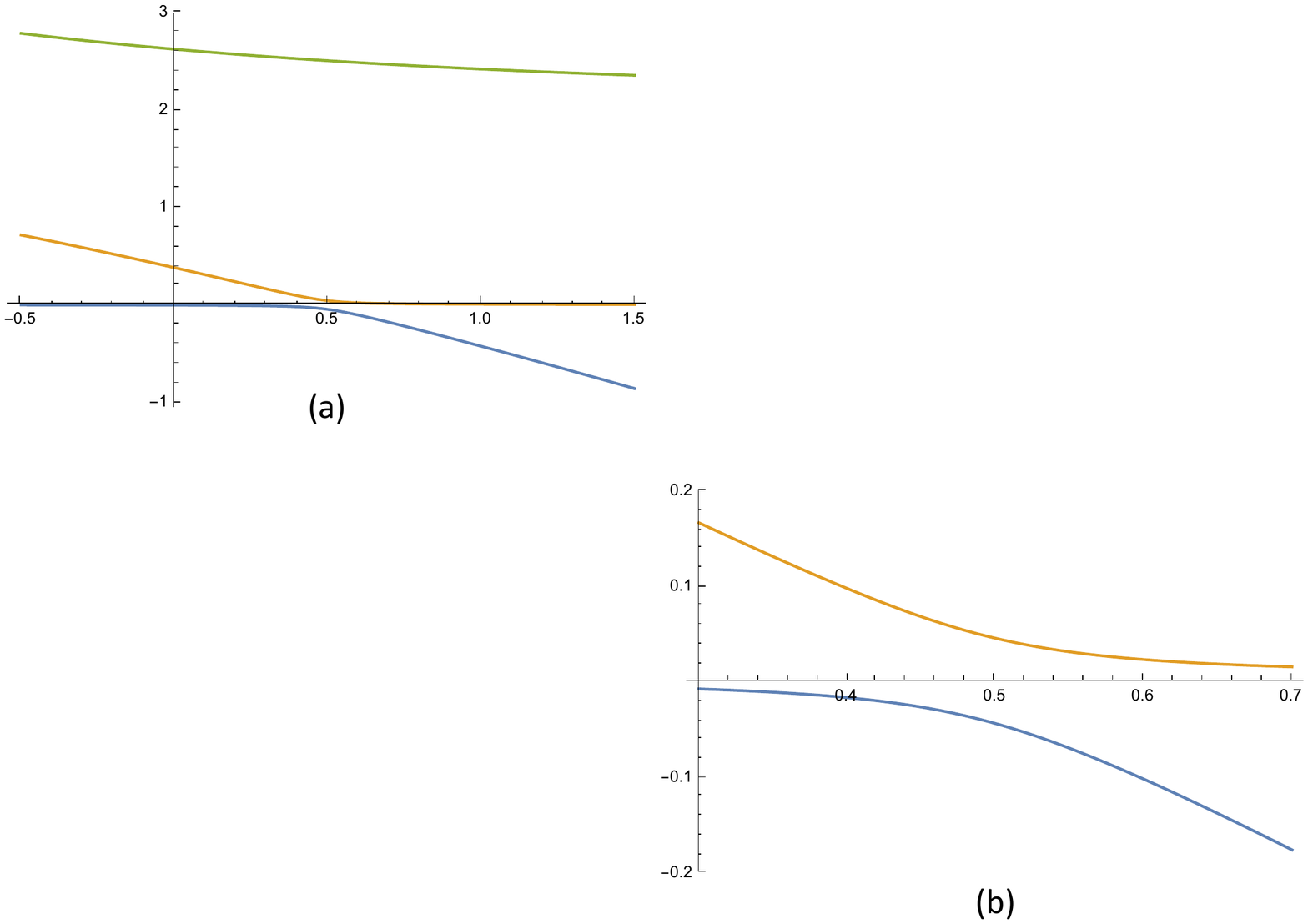}}
\caption{\small{$N_f=2$. (a) Evolution of the three eigenvalues as one varies $\mu^2/a$ for $b = 0.1$ and $\mu_2^2 = 2 \mu_1^2$. One of the three eigenvalue always lies much higher than the other two and is not much affected by the axion. (b) Blow up of the lower part of the figure showing the repulsion (and mixing) of the two lower eigenvalues. 
}}
\label{fig:fig8}
\end{figure}

\begin{figure}
\centerline{\includegraphics[scale=0.45,angle=0]{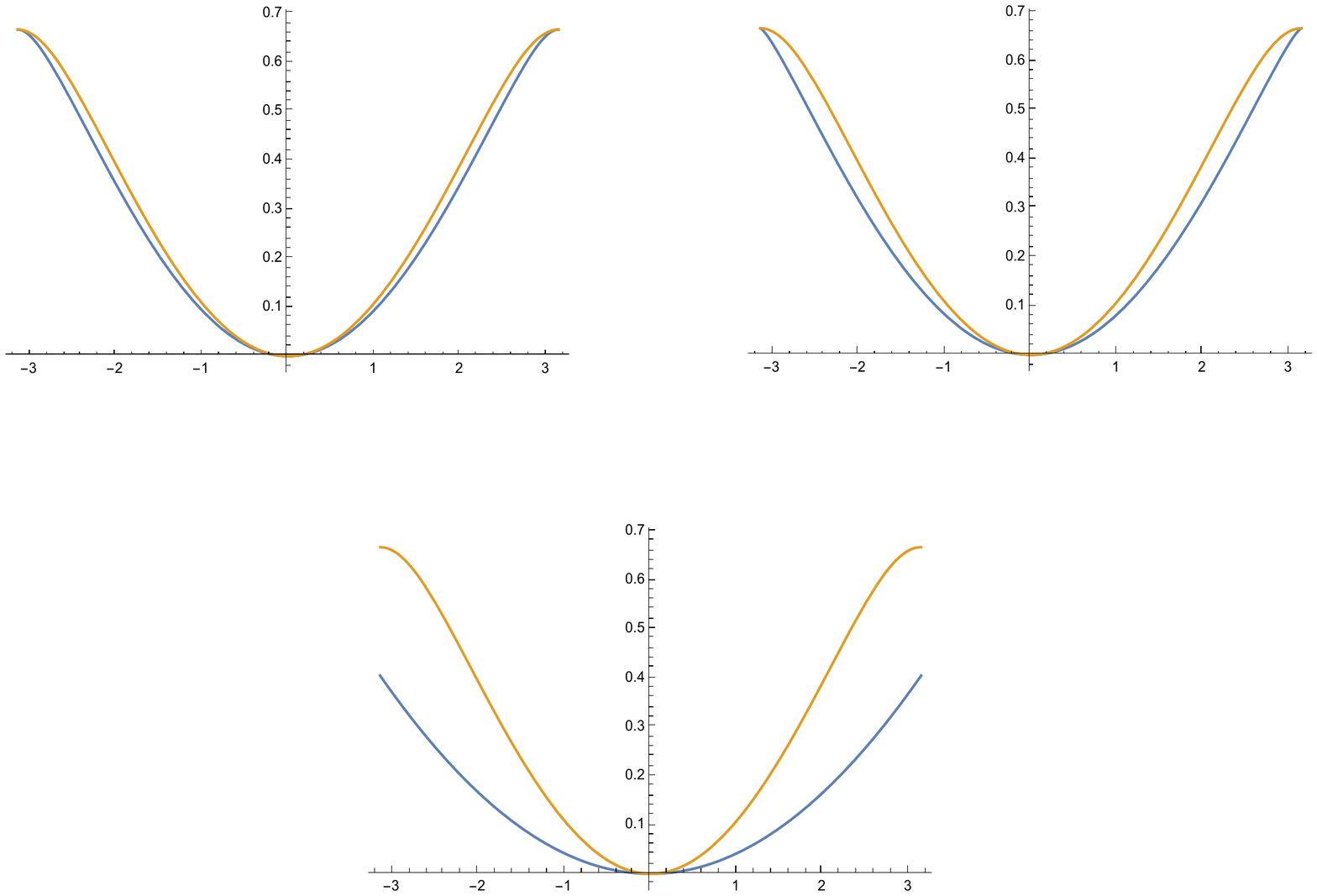}}
\vspace{-1cm}
\caption{\small{Comparing the conventional axion potential (yellow curves) with the ``exact" one (blue curves) for $N_f =2, \mu_d^2 = 2 \mu_u^2$ and at three  values of $\mu_u^2/a$: $0.25, 0.5~ {\rm (critical~value)},~ 2.5$. In the first two cases the two potentials (but not necessarily their derivatives) agree at $\zeta = \pm \pi$ while in the third (overcritical) case even the values of the potentials disagree at the boundary of the periodicity interval. 
}}
\label{fig:fig9}
\end{figure}

This is indeed fully supported by the numerical results shown in Figs.~\ref{fig:fig7} and~\ref{fig:fig8} for $N_f=1$ and $N_f=2$, respectively.
We have solved, using Mathematica, the minimization conditions at fixed $\zeta$ and reconstructed this way the axion potential (see Fig.~\ref{fig:fig9}). We then clearly see that, while at small $\mu_{u,d}^2/a$ the potential has a regular maximum  around $\zeta = \pi$ which coincides with the one of~(\ref{MINV5}) and agrees well with it elsewhere, as we increase $\mu_{u,d}^2/a$ above  the critical value $1- \mu_u^2/\mu_d^2$ (see Eq.~(\ref{cond})), 
the potential is lower that the one given by~(\ref{MINV5}) even at $\zeta = \pi$ and, by 
periodicity must develop a spike at that point. As we finally go much beyond the critical point, the true potential has nothing to do with the conventional one.

As in the $N_f=1$ case also here, the description of physics in terms of a single axion field is no longer appropriate when we are the vicinity of the condition~(\ref{cond}). In that case only one ``heavy" field can be integrated out and a  description in terms of two light fields is more appropriate. 

\section{Conclusions}
\label{conclusions}
\setcounter{equation}{0}

 The phase structure of QCD associated with spontaneous $CP$ breaking at $\theta=\pi$ may, potentially (depending on  parameters like quark masses and topological susceptibility, their ratios and temperature dependence), have important implications on the axion potential and it's cosmological ``phenomenology".

 In the present work we employed the effective chiral Lagrangian approach to investigate the inter-relation between spontaneous $CP$ breaking in QCD at  $\theta=\pi$ and the axion potential near the boundary of its periodicity interval. 
Formally, the effective Lagrangian approach is applicable at low energies and, in particular, when all mass parameters (notably quark masses) are small with respect to the QCD scale, $\Lambda$. We also look at the large-$N$ limit in which  we can have ratios of quark masses to $\Lambda$ small but still much larger then $1/N$. 
This allows us to identify and reliably investigate the existence, at $\theta=\pi$, of a second order phase transition point on the hypersurface dividing the region in parameters space where $CP$ is spontaneously broken from the one where it is not. 
The second order point is characterized by one of the PNGB mass going to zero and by the topological susceptibility (which can be seen as the order parameter) to diverge.

For generic masses the phase structure of QCD reveals a line of first order transitions, associated with spontaneous $CP$ breaking at $\theta=\pi$, along the negative real axis in the complex  $\mu_1^2 e^{i\theta}$ mass plane ($\mu_1$ being the lowest quark mass). The first order line extends all the way from $-\infty$ to the second order point without reaching the chiral point at the origin. The position of the second order transition depends on all other parameters (mass ratios and the susceptibility related parameter we called $a$). 
A similar phase structure is obtained by working in the complex quark-mass-determinant plane. 
 
It is the existence of this second order point which has the most dramatic effect on the axion potential.
Clearly, upon introducing the axionic field into the effective Lagrangian there is no more a $\theta$ dependence and no strong-$CP$ breaking. However, precisely around the point in parameter space (quark masses and topological susceptibility) where, in the absence of the axion, the condition for having a zero mass boson is met, we find large mixing between the would be massless particle and the axion. In this region one cannot integrate out all the PNGB since one of them becomes very light with a mass of the order of the axion mass. Hence, in this region, the notion of an axionic potential which depends on just the axion field (obtained upon integration out all the PNGB) is not viable and should be replaced by a potential which depends on the two above mentioned light degrees of freedom as discussed in Sect.~\ref{axion}. This potential is obtained upon integrating out all the other much heavier PNGBs.

\begin{figure}[htbp]
\centerline{\includegraphics[scale=0.70,angle=0]{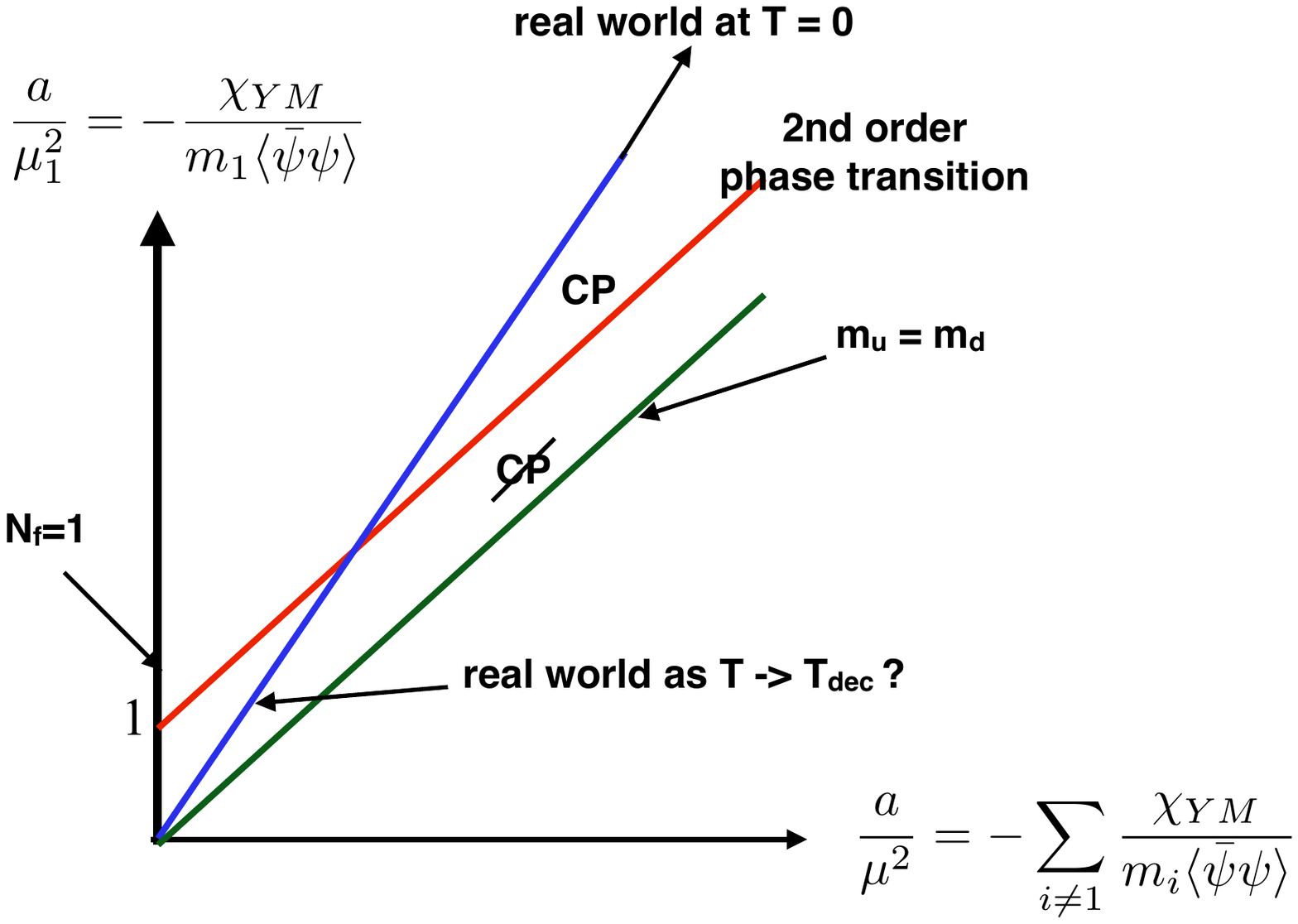}}
\vspace{-3.cm}
\caption{\small{Phase diagram for $N_f =2$. The red line separates the phases with broken and unbroken $CP$ at $\theta = \pi$ and corresponds to a second-order transition with a massless particle. For $m_d >> m_u$ we recover the $N_f=1$ case represented by the vertical axis. Also shown is the $m_u = m_d$ case lying entirely in the $CP$ broken phase. The real world at $T=0$  is far up on the blue line (representing $m_d \sim 2m_u$). As $T$ is increased towards $T_{dec}$ the real world will stay on the blue line (since $m_d /m_u$  is $T$-independent) but  may move down and cross the red line as indicated in the picture. Present lattice data seem to disfavour this possibility.}}
\label{fig:fig10}
\end{figure}

Given the actual physical numerical values of the parameters (for $N_f=2$ and $N_f=3$) we see that, at zero temperature, we are not in the region of the parameter space where the concept of an axion potential and the derived result for the axion mass should be modified. However, if, as we raise the temperature while staying below the deconfinement transition (which for QCD is not a sharp transition), the corresponding YM topological susceptibility (and hence the parameter $a$) drops faster with the temperature than the quark condensate so as to allow $\mu^2/a$  to increase by about an order of magnitude, we will  enter into this intriguing region (see Fig.~\ref{fig:fig10}). 

It seems, however, that lattice calculations (see e.g.~\cite{Berkowitz:2015aua,Borsanyi:2015cka,Bonati:2013tt} as well as~\cite{Chen:1998xw,Edwards:1999zm}) show a rather mild $T$-dependence of both $\chi_{YM}$ and the quenched chiral condensate with a sharp drop (but not necessarily vanishing) of both above a similar value of $T$. There does not seem to be a clean window in which $\mu^2/a$ increases by the above-mentioned order of magnitude. It would be desirable to have detailed lattice data on both $\chi_{YM}$ and the  planar
chiral condensate by a single group using the same Montecarlo configurations. It would be particularly interesting to study the pure number $\chi_{YM}/\langle m \bar{\psi} \psi \rangle$ in the vicinity of the above-mentioned drop and also check its $N$-dependence (expected to be $1/N$).

An obviously related issue is whether there is a critical temperature $T_{top}$ above which $\chi_{YM}$ vanishes, at least in the large-$N$ limit (dilute instantons~\cite{Gross:1980br,Schafer:1996wv,Ringwald:1999ze}, for instance, predict $\chi_{YM} \sim e^{-c N}$) and, in that case, whether $T_{top}$ can be higher than $T_{ch}$, the temperature above which chiral symmetry is restored. Under reasonable assumptions, claims that $\chi_{YM}$ should vanish above $T_{ch}$ were made in the past~\cite{Veneziano:1980xs,Cohen:1996ng} leaving open the possibility that $\chi_{YM}$ goes to zero either together or before $\langle \bar{\psi} \psi \rangle$ does it.
 
Although some old lattice calculations~\cite{DiGiacomo:1991qm} appear to point in the opposite direction (and such a possibility has its own effective Lagrangian formulation~\cite{Meggiolaro:1992wc}), more recent simulations of the pure gauge theory~\cite{Alles:1996nm,DelDebbio:2004vxo} suggest the existence of a similar (or even identical) value for the temperatures of deconfinement, chiral restoration and $U_A(1)$ restoration. Above the transition temperature the dilute instanton gas approximation seems to set in. Actually there is lattice evidence~\cite{Lucini:2004yh} that $\chi_{YM}$ drops rather fast above $T_c$ for large $N$ (and even at $N=3$ a substantial decrease of $\chi_{YM}$ is visible~\cite{Gattringer:2002mr,Bornyakov:2013iva,Xiong:2015dya}) and may actually go to zero above it for $N \to \infty$. However, it is not clear what the ratio $\langle \bar \psi \psi\rangle_{planar}/\chi_{YM}$ does around $T_c$. It would thus be very interesting to plan new lattice projects dedicated to the calculation of $\chi_{YM}$ and $\langle\bar \psi \psi\rangle$ in the planar limit across the phase transition.

Recently, using the mixed $CP$/Center discrete anomaly matching (together with some other plausible assumptions), it was shown~\cite{GKKS} that in YM theory the $CP$ symmetry is spontaneously broken at $\theta=\pi$ and zero temperature  and that the temperature $T_{res}$ at which $CP$ is restored is higher than the deconfinement temperature, i.e.\ $T_{res} \geq T_{dec}$. This result seems to be going in favor of the scenario advocated in~\cite{Veneziano:1980xs,Cohen:1996ng}. Breaking of $CP$ in YM connects smoothly with $CP$-breaking in, say,  $N_f=1$ QCD  at $\mu^2/a >1$. As we increase the temperature, if $CP$ were restored before reaching $T_{dec}$, it would suggest that, in its QCD analog, $\mu^2/a$ would go down till, at $T_{res}$, it reaches 1, which is precisely the opposite of what we were advocating, i.e.\ a ratio $\mu^2/a$ increasing with temperature. Hence the statement $T_{res} \geq T_{dec}$ is an (admittedly very mild) indication in favor of the scenario in which the finite temperature axion potential has to be revised in a certain range of temperature. Even if such a revision would be necessary, it remains to be seen whether it would make any substantial difference with respect to the standard calculations~\cite{Preskill:1982cy} (see also~\cite{AS}, \cite{DF}) of axionic dark matter abundance.

\vspace{.7cm}
\noindent {\large \textbf{Acknowledgements} }

\vspace{.2cm}
We thank D. Gaiotto, Z. Komargodski  and N. Seiberg for informing us of their work~\cite{GKS} prior to posting it. We also thank Z. Komargodski  for useful comments on a preliminary version of this manuscript as well as M. D'Elia, L. Giusti and  E. Vicari  for discussions about lattice results on Yang-Mills and quenched QCD at finite temperature. S.Y.\ would like to thank O. Aharony and M. Peskin for discussions. G.V.\ wishes to acknowledge an illuminating discussion with M. Shifman. The work of S.Y.\ is supported in part by the I-CORE program of the Planing and Budgeting Committee (grant number 1937/12), the US-Israel Binational Science Foundation (BSF), the Israel-Germany Foundation (GIF) and the ISF Center of Excellence.  

\begin{appendix}

\section{Ward--Takahashi identities}
\label{appA}
\setcounter{equation}{0}

In this Appendix we derive the WTIs for the anomalous $U_A(1)$ currents in
QCD and check that the two-point amplitudes derived from the 
effective Lagrangian in Sect.~\ref{effective} exactly satisfy them. We start from the anomaly equation in~(\ref{Q+anomaly}), but written for a single flavor
\begin{eqnarray}
\partial_\mu J^{\mu}_{5i} = 2 Q + 2 m_i P_i~~;~~J_{5i}^{\mu} = {\bar{\psi}}_i \gamma^\mu \gamma_5 \psi_i ~~;~~P_i = i{\bar{\psi}}_i \gamma_5 \psi_i\, .
\label{anoma1fla}
\end{eqnarray}
Inserting the previous anomaly equation in a two-point amplitudes with another operator $O(y)$ we get
\begin{eqnarray}
\partial_\mu \langle J^\mu_{5i} O(y) \rangle = \langle 2Q(x) O(y) \rangle + \delta (x^0 - y^0)
\langle [ J_{5i}^0 , O(y) ] \rangle + \langle 2 m_i P_i (x) O(y)\rangle \, ,
\label{2poi}
\end{eqnarray}
that in Fourier space, after a partial integration,  becomes
\begin{eqnarray}
&& \int d^4 x \,\, {\rm e}^{ipx} \langle 2Q(x) O(y) \rangle + \langle [ Q_{5i} , O(y) ] 
+ \int d^4 x \,\, {\rm e}^{ipx}  \langle 2 m_i P_i (x) O(y)\rangle  \nonumber \\
&& = -i \int d^4 x \,\, {\rm e}^{ipx}  \langle p_\mu J^{\mu}_{5i} (x) O (y) \rangle ~~~;~~~
i =1, \dots, N_f \, ,
\label{WIO}
\end{eqnarray}
where $Q_{5i} = \int d^3 x \,\, J^0_{5i} (x)$. For $O(y) = Q(y)$ the second term does not contribute and we get
\begin{eqnarray}
\hspace{-1.cm}\int d^4 x \,\, {\rm e}^{ipx} \langle 2Q(x) Q(y) \rangle 
+ \int d^4 x \,\, {\rm e}^{ipx}  \langle 2 m_i P_i (x) Q(y)\rangle =
 -i \int d^4 x \,\, {\rm e}^{ipx}  \langle p_\mu J^{\mu}_{5i} (x) Q (y) \rangle \, ,
\label{Oy=Qy}
\end{eqnarray}
while, for $O(y) = 2 m_j P_j (y)$, the commutator gives  $[Q_{5i}, P_j ]= -2i {\bar{\psi}}_i \psi_i \delta_{ij}$ 
and we get
\begin{eqnarray}
\hspace{-1.cm}&&\int d^4 x \,\, {\rm e}^{ipx} \langle 2Q(x)  2m_j P_j (y) \rangle + 2i \mu_i^2 F_\pi^2 + \int d^4 x \,\, {\rm e}^{ipx}  \langle 2 m_i P_i (x) 2 m_j P_j (y)\rangle
  \nonumber \\
\hspace{-1.cm}&& =
   -i \int d^4 x \,\, {\rm e}^{ipx}  \langle p_\mu J^{\mu}_{5i} (x)  2 m_j P_j  (y) \rangle \, ,
\label{Oy=2mP}
\end{eqnarray}
having made use of the Gell-Mann--Oakes--Renner relation $-2 \delta_{ij} m_i \langle {\bar{\psi}}_i \psi_i \rangle = \delta_{ij} \mu_i^2 F_\pi^2$. 

One checks that the following two-point amplitudes satisfy the previous anomalous WTIs and we get
\begin{eqnarray}
\int d^4 x {\rm e}^{i p x}\langle Q(x) Q(y) \rangle =i \frac{aF_\pi^2}{2} 
 \prod_{i=1}^{N_f} 
\frac{p^2 - \mu_i^2}{p^2 - M_i^2} = i \frac{aF_\pi^2}{2} \left[ 1 - a \sum_{i=1}^{N_f} 
\frac{1}{p^2 - \mu_i^2} \right]
^{-1} \, ,
\label{qq}
\end{eqnarray}
\begin{eqnarray}
\int d^4 x {\rm e}^{i p x}\langle Q(x) 2 m_i P_i \rangle =i \frac{2 \mu_i^2}{p^2- \mu_i^2} \frac{aF_\pi^2}{2}
 \prod_{j=1}^{N_f} 
\frac{p^2 - \mu_j^2}{p^2 - M_j^2} \, ,
\label{qq mPi}
\end{eqnarray}
\begin{eqnarray}
\int d^4 x {\rm e}^{i p x}\langle J_{5\mu}^{(i)}(x) Q(y) \rangle = -\frac{2  p_{\mu}}{p^2- 
\mu_i^2} \frac{aF_\pi^2}{2} 
 \prod_{j=1}^{N_f} 
\frac{p^2 - \mu_j^2}{p^2 - M_j^2} \, ,
\label{Jq}
\end{eqnarray}
\begin{eqnarray}
&& \int d^4 x {\rm e}^{i p x}\langle  2 m_i P_i (x) 2 m_j P_j \rangle =\nonumber \\
&&=i\frac{2 F_\pi^2 \mu_i^4}{p^2- \mu_i^2} \delta_{ij} 
+i \frac{4 \mu_i^2 \mu_j^2}{(p^2- \mu_i^2)(p^2- \mu_j^2)} \frac{aF_\pi^2}{2}
 \prod_{k=1}^{N_f} 
\frac{p^2 - \mu_k^2}{p^2 - M_k^2} \, ,
\label{mPimPj}
\end{eqnarray}
\begin{eqnarray}
&& \int d^4 x {\rm e}^{i p x}\langle J_{5\mu}^{(i)}(x)  2 m_j P_j \rangle =\nonumber \\
&&=  - \frac{2  F_\pi^2 \mu_i^2 p_{\mu} }{p^2- \mu_i^2} \delta_{ij}
-  \frac{4   p_{\mu} \mu_j^2}{(p^2- \mu_i^2)(p^2- \mu_j^2)} \frac{aF_\pi^2}{2}
 \prod_{k=1}^{N_f} 
\frac{p^2 - \mu_k^2}{p^2 - M_k^2} \, .
\label{JmP}
\end{eqnarray}
Furthermore, the poles at $p^2 = \mu_i^2$ apparently present in~(\ref{mPimPj}) and~(\ref{JmP}) can be shown to be absent. The only poles present are at $p^2 = M_i^2$ and correspond to the masses of the physical mesons. The previous two-point amplitudes reproduce those in Sect.~\ref{effective} with the identification
\begin{eqnarray}
m_i P_i  \Longrightarrow \frac{F_\pi}{\sqrt{2}} \mu_i^2 v_i \, .
\label{identif}
\end{eqnarray}

\end{appendix}


\begin{thebibliography}{99}

\bibitem{RD}
R.~F. Dashen, Phys.\ Rev.\ {\bf D3} (1971) 1879.

\bibitem{JN}
J. Nuyts, Phys. Rev. Lett. {\bf 26} (1971) 1604; {\bf 27} (1971) 361.

\bibitem{MB}
M.~A.~B. B\'eg, Phys.\ Rev.\ {\bf D4} (1971) 3810.

\bibitem{SWCP}
S. Weinberg, Phys. Rev. Lett. {\bf 31} (1973) 494 and Phys. Rev. {\bf  D8} (1973) 4482.

\bibitem{tHooft}
G. 't Hooft, Phys. Rev. Lett. {\bf 37} (1976) 8 and Phys. Rev. {\bf D14} (1976) 3432.

\bibitem{RC}
R. Crewther, NATO Ad. Study Inst. Ser. B Phys. {\bf 55} (1980) 529. 

\bibitem{EW1}
E. Witten, Nucl. Phys. {\bf B156} (1979) 269.

\bibitem{GV}
G. Veneziano, Nucl. Phys. {\bf B159} (1979) 213.

\bibitem{DiVecchia:1981aev}
P.~Di Vecchia, K.~Fabricius, G.~C.~Rossi and G.~Veneziano, Nucl.\ Phys.\ {\bf B192} (1981) 392.
 
\bibitem{Giusti:2001xh}
  L.~Giusti, G.~C.~Rossi, M.~Testa and G.~Veneziano,
  Nucl.\ Phys.\  {\bf B628} (2002) 234.
 
\bibitem{DelDebbio:2004ns}
  L.~Del Debbio, L.~Giusti and C.~Pica, Phys.\ Rev.\ Lett.\  {\bf 94} (2005) 032003.
 
\bibitem{Ce:2015qha}
M.~C\'e, C.~Consonni, G.~P.~Engel and L.~Giusti, Phys.\ Rev.\ {\bf D92} (2015) 074502.
 
\bibitem{PDV1}
P. Di Vecchia, Phys. Lett. {\bf 85B} (1979) 357.

\bibitem{VB}
V. Baluni, Phys. Rev. {\bf D19} (1979) 2227.

\bibitem{Crewther:1979pi}
  R.~J.~Crewther, P.~Di Vecchia, G.~Veneziano and E.~Witten,
  Phys.\ Lett.\  {\bf 88B} (1979) 123.
   Erratum: [Phys.\ Lett.\  {\bf 91B} (1980) 487].

\bibitem{RST}
C. Rosenzweig, J. Schechter and C. Trahern, Phys. Rev. {\bf D21} (1980) 3388.

\bibitem{DVV}
P. Di Vecchia and G. Veneziano, Nucl. Phys. {\bf B171} (1980) 253.

\bibitem{NA}
P. Nath and R. Arnowitt, Phys. Rev. {\bf D23} (1981) 473. 

\bibitem{EW}
E. Witten, Annals of Physics {\bf 128} (1980) 363.

\bibitem{KO}
K. Kawarabayashi and N. Ohta, Nucl. Phys. {\bf B175} (1980) 477 and Prog. Theor. {\bf 66} (1981) 1789.

\bibitem{NO}
N. Ohta, Prog. Theor. Phys. {\bf 66} (1981) 1408.

\bibitem{PDV}
P. Di Vecchia, Acta Physica Austriaca, Suppl. XXII (1980) 341-381.

\bibitem{DVS}
P. Di Vecchia and F. Sannino, Eur. Phys. J. Plus, {\bf 129} (2014) 262.

\bibitem{Smilga:1998dh}
  A.~V.~Smilga,
  Phys.\ Rev.\  {\bf D59} (1999) 114021.

\bibitem{MC1}
M. Creutz, Phys. Rev. Lett. {\bf 92} (2004) 201601. 

\bibitem{MC2}
M. Creutz, Ann. Phys. {\bf 322} (2007) 1518, arXiv:hep-th/0609187; hep-th/0303018.

\bibitem{GKKS}
D. Gaiotto, A. Kapustin, Z. Komargodski and N. Seiberg, JHEP {\bf 1705} (2017) 091. 

\bibitem{Nati} 
N. Seiberg, talk given at STRINGS 2017.

\bibitem{GKS} 
D. Gaiotto, Z. Komargodski and N. Seiberg, arXiv:hep-th/1708.06806.

\bibitem{diCortona:2015ldu}
  G.~Grilli di Cortona, E.~Hardy, J.~Pardo Vega and G.~Villadoro,
  JHEP {\bf 1601} (2016) 034.

\bibitem{Komargodski:2017dmc}
Z. Komargodski, A. Sharon, R. Thomgren and X. Zhou, arxiv:1705.0478 [hep-th].


\bibitem{Sharpe:1992ft}
S.R. Sharpe, Phys. Rev. {\bf D46} (1992) 3146, hep-lat/9505020.

\bibitem{Bernard:1992mk}
C.W. Bernard and M.F. Golterman, Phys. Rev.{\bf D46} (1992) 853, hep-lat/9204007.


\bibitem{Giusti:2002rx}
L. Giusti, Nucl. Phys. Proc. Suppl. {\bf 119} (2003) 149,  hep-lat/0211009.

\bibitem{SWax}
S. Weinberg,  Phys. Rev. Lett. {\bf 40} (1978) 223. 

\bibitem{FW}
F. Wilczek, Phys. Rev. Lett {\bf 40} (1978) 279.

\bibitem{PQ}
R. Peccei and H. Quinn, Phys. Rev. Lett. {\bf 38} (1977) 1440 and 
Phys. Rev. {\bf D16} (1977)1791.

\bibitem{BT}
W.~A.~Bardeen and H.~H.~Tye, Phys. Lett. {\bf 74B} (1978) 229.

\bibitem{Preskill:1982cy}
J.~Preskill, M.~B.~Wise and F.~Wilczek, Phys. Lett. {\bf 120B} (1983) 127.

\bibitem{Berkowitz:2015aua}
 E.~Berkowitz, M.~I.~Buchoff and E.~Rinaldi, Phys.\ Rev.\ {\bf D92} (2015) 034507.

\bibitem{Borsanyi:2015cka}
 S.~Borsanyi {\it et al.},  Phys.\ Lett.\ {\bf 752B} (2016) 175.
 
\bibitem{Bonati:2013tt} 
  C.~Bonati, M.~D'Elia, H.~Panagopoulos and E.~Vicari,
  Phys.\ Rev.\ Lett.\  {\bf 110}, no. 25, 252003 (2013).
  
\bibitem{Chen:1998xw} 
  P.~Chen {\it et al.},
  In *Vancouver 1998, High energy physics, vol. 2* 1802-1808 [hep-lat/9812011].

\bibitem{Edwards:1999zm} 
  R.~G.~Edwards, U.~M.~Heller, J.~E.~Kiskis and R.~Narayanan,
  Phys.\ Rev.\ D {\bf 61}, 074504 (2000).
  
\bibitem{Gross:1980br}
  D.~J.~Gross, R.~D.~Pisarski and L.~G.~Yaffe,
  Rev.\ Mod.\ Phys.\  {\bf 53} (1981) 43.

\bibitem{Schafer:1996wv}
  T.~Sch\"afer and E.~V.~Shuryak,
  Rev.\ Mod.\ Phys.\  {\bf 70} (1998) 323.
  
\bibitem{Ringwald:1999ze}
  A.~Ringwald and F.~Schrempp,
  Phys.\ Lett.\ B {\bf 459} (1999) 249:
  
\bibitem{Veneziano:1980xs}
  G.~Veneziano,
  Phys.\ Lett.\  {\bf 95B} (1980) 90.

\bibitem{Cohen:1996ng}
  T.~D.~Cohen,
  Phys.\ Rev.\ D {\bf 54} (1996) R1867.

\bibitem{DiGiacomo:1991qm}
  A.~Di Giacomo, E.~Meggiolaro and H.~Panagopoulos,
  Phys.\ Lett.\ B {\bf 277} (1992) 491.

\bibitem{Meggiolaro:1992wc}
  E.~Meggiolaro,
  Z.\ Phys.\ C {\bf 62} (1994) 669 and  Z.\ Phys.\ C {\bf 62} (1994) 679.

\bibitem{Alles:1996nm}
  B.~Alles, M.~D'Elia and A.~Di Giacomo,
  Nucl.\ Phys.\ B {\bf 494} (1997) 281, Erratum: [Nucl.\ Phys.\ B {\bf 679} (2004) 397].

\bibitem{DelDebbio:2004vxo}
  L.~Del Debbio, H.~Panagopoulos and E.~Vicari,
  JHEP {\bf 0409} (2004) 028.

\bibitem{Lucini:2004yh}
  B.~Lucini, M.~Teper and U.~Wenger,
  Nucl.\ Phys.\ B {\bf 715} (2005) 461.

\bibitem{Gattringer:2002mr}
  C.~Gattringer, R.~Hoffmann and S.~Schaefer,
  Phys.\ Lett.\ B {\bf 535} (2002) 358.

\bibitem{Bornyakov:2013iva}
  V.~G.~Bornyakov, E.-M.~Ilgenfritz, B.~V.~Martemyanov, V.~K.~Mitrjushkin and M.~MŸller-Preussker,
  Phys.\ Rev.\ D {\bf 87} (2013) no.11, 114508.
 
\bibitem{Xiong:2015dya}
  G.~Y.~Xiong, J.~B.~Zhang, Y.~Chen, C.~Liu, Y.~B.~Liu and J.~P.~Ma,
  Phys.\ Lett.\ B {\bf 752} (2016) 34.
 
 \bibitem{AS} 
 L.~F.~ Abbott and P.~Sikivie, Phys. Lett. {\bf 120B} (1983) 133.
 
 \bibitem {DF} 
 M.~Dine and W.~Fischler, Phys. Lett. {\bf 120B} (1983) 137.
 
\end{thebibliography}
\end{document}